\documentclass[
aip,
 amsmath,amssymb,
 reprint,
 nofootinbib,
]{revtex4-1}
\usepackage{amsmath}
\usepackage{amssymb}
\usepackage{color}
\usepackage{graphicx}
\usepackage{tabularx}
\usepackage{dcolumn} 
\usepackage{bm}      
\usepackage{array}   
\usepackage{amsfonts}
\usepackage[bookmarks=false,colorlinks,citecolor=red]{hyperref}

\include{MyCommand}

\begin{document}

\title{Fractional Path Integral Monte Carlo}

\author{Mamikon Gulian}
 \affiliation{Department of Applied Mathematics, Brown University, Providence, RI 02912}

\author{Haobo Yang}
\affiliation{
Department of Chemistry, Brown University, Providence, RI 02912
}
\author{Brenda M. Rubenstein}
\affiliation{
Department of Chemistry, Brown University, Providence, RI 02912
}

\date{\today}

\begin{abstract}
Fractional derivatives are nonlocal differential operators of real order that often appear in 
models of anomalous diffusion and a variety of nonlocal phenomena. Recently, a version of the Schr\"odinger Equation containing a fractional Laplacian has been proposed. In this work, we develop a Fractional Path Integral Monte Carlo algorithm that can be used to study the finite temperature behavior of the time-independent Fractional Schr\"odinger Equation for a variety of potentials. In so doing, we derive an analytic form for the finite temperature fractional free particle density matrix and demonstrate how it can be sampled to acquire new sets of particle positions. We employ this algorithm to simulate both the free particle and $^{4}$He (Aziz) Hamiltonians. We find that the fractional Laplacian strongly encourages particle delocalization, even in the presence of interactions, suggesting that fractional Hamiltonians may manifest atypical forms of condensation. Our work opens the door to studying fractional Hamiltonians with arbitrarily complex potentials that escape analytical solutions.     
\end{abstract}


\maketitle

\section{\label{sec:Introduction} Introduction}

Fractional calculus is the study and use of differential operators of real order. Such operators are nonlocal except for the case of nonnegative integer order. Space- and time-fractional differential equations have emerged as the correct way to model anomalous diffusion, \cite{Klafter_Physics_World, Meerschaert_Stochastic} nonlocal phenomena,\cite{Lazopoulos_MRC} memory effects,\cite{Du_Nature} and many non-equilibrium \cite{Vlahos_Tutorial} and turbulent\cite{Majda_PR, Richardson_PRSA} systems. 

Anomalous diffusion is any diffusion in which the mean square displacement is not linear in time, i.e., 
\begin{equation}\label{E:mean_square_displacement}
\langle x^2 \rangle \sim t^{\frac{2}{\alpha}},
\end{equation}
where ${\alpha \neq 2}$, in contrast to the ${\alpha = 2}$ case, which corresponds to normal/Brownian diffusion. If ${\alpha < 2}$, the diffusion is superdiffusive, while if ${\alpha > 2}$, the diffusion is subdiffusive. One can think of superdiffusion for $1<\alpha<2$ as a regime between normal diffusion and ballistic motion (for which ${\alpha = 1}$, implying that the average displacement is proportional to time). A simple example of superdiffusion is a plasma whose particles would ordinarily be expected to undergo diffusion because of their fluid nature, but instead exhibit superdiffusive behavior because of the presence of a magnetic field that favors ballistic motion.\cite{Corkum_PRL, Vlahos_Tutorial} On the other hand, traps (such as in porous media\cite{Benson_Transport}) or formations of eddies in a fluid\cite{Vlahos_Tutorial, Solomon_Physica_D, Weeks_Physica_D} may introduce waiting times that slow down the fluid's particles, pushing their motion toward subdiffusion. In general, both tendencies can occur, resulting in a unique process with ``competition'' between super- and sub-diffusive overall behavior. 

Anomalous diffusion differs significantly from normal diffusion, in more ways than can be expected from the scaling relation given by Relation \eqref{E:mean_square_displacement}. Anomalous diffusion is often characterized by heavy tails, infinite variance, and inhomogeneity in space and time. These are characteristics of Continuous Time Random Walks (CTRWs), \color{black}a broad family of anomalous diffusions which model both waitings times and long jumps. \cite{Meerschaert_Scheffler, Meerschaert_Stochastic} \color{black} The simplest subfamily of CTRWs is isotropic ${\alpha}$-stable Levy motion.

Just as the standard Laplacian ${(-\Delta)}$ in ${\mathbb{R}^n}$ is the generator of Brownian motion, a random walk in which the jumps can be drawn from a normal distribution, for ${0 < \alpha < 2}$, the fractional Laplacian ${(\Delta)^{\alpha/2} := -(-\Delta)^{\alpha/2}}$ in ${\mathbb{R}^n}$ is the generator of isotropic ${\alpha}$-stable Levy motion.\cite{Meerschaert_Stochastic} In this motion, jumps are drawn from an ${\alpha}$-stable distribution, which exhibits power-law tails. The isotropic ${\alpha}$-stable distribution is defined by a position parameter and a scale parameter, which for the normal ${\alpha = 2}$ case, reduces to the mean and variance, respectively. 

\begin{figure}[h]
  \centering
   \includegraphics[width=\columnwidth]{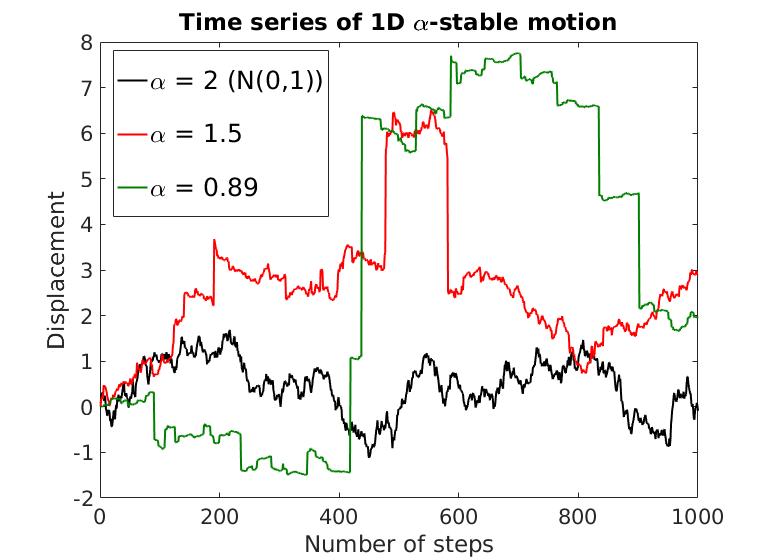}
  \caption{Illustration of superdiffusive, $\alpha$-stable random walks and comparison with the normal, $\alpha=2$, case in one dimension. The position parameter is set to $0$ and the scale parameter is set to $1/\sqrt{2}$, giving a variance of $1$ when $\alpha=2$. For $0 < \alpha < 2$, the motion is discontinuous in time and has infinite variance; for $\alpha < 1$, the mean itself does not exist. }
\end{figure}

\color{black}The field of physics where anomalous diffusion is most widespread is perhaps plasma physics, where anomalous diffusion has been observed for decades.\cite{Yoshikawa_POF, Geissler_PR} \color{black} This has prompted theoretical discussions on how the equations of magnetohydrodynamics can drive anomalous diffusion\cite{Drummond_POF, Corkum_PRL, Zaslavsky_PR} and applications of CTRWs to  plasma physics.\cite{Balescu_PRE} Anomalous diffusion has moreover been observed\cite{Kapitulnik_PRL} in percolation clusters, which has since been explained theoretically. \cite{Gefen_PRL} Inspired by this finding, the role of anomalous diffusion in disordered superconductors (and other disordered media) has also been elucidated.\cite{Bouchaud_PR} 

\renewcommand{\thefootnote}{$\star$} 
Recently, a nonlinear fractional Schr\"odinger equation has been rigorously derived from a mathematical physics perspective as the continuum limit of a system of discrete nonlinear Schr\"odinger equations with increasingly long-range interactions.\cite{Kirkpatrick_CMP} Laskin\cite{Laskin_1, Laskin_2, Laskin_3, Laskin_4} first developed fractional quantum mechanics, following the early considerations of Kac.\cite{Kac_Berkeley} Starting from the Feynman path-integral formulation of quantum mechanics, Laskin replaced integrals over Brownian paths with integrals over $\alpha$-stable Levy paths.  He showed that this amounts to replacing the standard Laplacian by the
(Riesz)\footnote{
We distinguish the Riesz fractional Laplacian from other forms because there are several ways to define the fractional Laplacian. Although all of the popular fractional Laplacians agree with each other in ${\mathbb{R}^n}$ , they no longer agree on a bounded domain.\cite{Hermann_Fractional} The use of the Riesz Laplacian $\Delta^{\alpha/2}:=-(-\Delta)^{\alpha/2}$, which can be characterized in ${\mathbb{R}^n}$ as having Fourier symbol \color{black}-$|\vec{p}|^\alpha$\color{black},  is required by the derivation of fractional quantum mechanics.
                      } 
fractional Laplacian in the Schr\"odinger Equation, leading to
\begin{equation}\label{E:FSE}
i\hbar \frac{\partial \psi(\vec{r}, t)}{\partial t} = - \hbar D_\alpha \Delta^{\alpha/2}\psi(\vec{r}, t) 
+ V(\vec{r})\psi(\vec{r}, t). 
\end{equation}
The equation is called the space-fractional Schr\"odinger equation. 
\color{black}
Laskin showed that this equation enjoys the same basic properties that allow the usual Schr\"odinger equation to serve as a foundation for quantum mechanics; for example, since
$-\Delta^{\alpha/2}$ is self-adjoint, observables are real valued.\cite{Laskin_4} \color{black}
To date, solutions to the fractional Schr\"odinger Equation have been derived for a variety of analytically tractable Hamiltonians, including the free particle, square well, and particle in a box Hamiltonians.\cite{Guo_Fractional,Dong_Fractional,Secchi_Fractional,Felmer_Fractional} Solutions to Hamiltonians containing more realistic potentials have so far been few and far between, limiting the study of potentially rich fractional mathematics and physics. 

In this paper, we introduce a new fractional Path Integral Monte Carlo method capable of modeling the finite temperature physics of a wide variety of space-fractional Hamiltonians. The method computes finite temperature observables using the Path Integral Monte Carlo (PIMC) algorithm popularly used to characterize boson and Boltzmannon phase diagrams \cite{Ceperley_Rev_Mod_Phys,Ceperley_Pollock_PRL,Pollock_Ceperley_PRB} to sample the many body fractional partition function. Key to being able to use the PIMC algorithm in this context is being able to express \emph{and sample} the free finite temperature fractional density matrix. We therefore derive an analytical form for this fractional density matrix and show how it can be both computed and sampled. 

\renewcommand{\thefootnote}{$\dagger$} 
Interestingly, we find that the probability distribution function that represents the fractional density matrix manifests dramatically different asymptotic behaviors depending upon the fractional exponent present in the Hamiltonian. Whereas for fractional exponent $\alpha=2$ the distribution reduces to the usual Gaussian case, for $\alpha<2$, the distribution possesses heavy tails, while for $\alpha>2$, the distribution becomes negative. Setting aside the $\alpha>2$ case, which is typically excluded\footnote{There is no $\alpha$-stable process for $\alpha > 2$.} in discussions of the fractional laplacian, in the remainder of this work, we illustrate how our algorithm may be used to compute finite temperature observables including energies and radii of gyration for two illustrative Hamiltonians containing $\alpha \le 2$ fractional Laplacians: 1) the free particle Hamiltonian; and 2) the $^{4}$He Hamiltonian using the Aziz\cite{Aziz_Helium} potential. We end with a discussion of the general applicability of our method and the intriguing fractional physics it may reveal.    

\section{\label{sec:Methods} Theory and Algorithms}

\subsection{\label{sec:Review_PIMC} Overview of the Path Integral Monte Carlo Method}

Since our method may be viewed as a generalization of the conventional Path Integral Monte Carlo algorithm, we begin this section with a review of PIMC. As described in significantly more detail by Ceperley,\cite{Ceperley_Rev_Mod_Phys} the finite temperature equilibrium expectation values of a system observable, $\hat{O}$, in the position basis may be expressed as 

\begin{equation}
\langle \hat{O} \rangle = Z^{-1} \int d\vec{R} d\vec{R}^{'} \rho(\vec{R},\vec{R}^{'}; \beta) \langle \vec{R} | \hat{O} | \vec{R}^{'} \rangle,
\label{Observable}
\end{equation}
where $Z$ denotes the system's partition function 

\begin{equation}
Z = \int d\vec{R} \rho(\vec{R}, \vec{R}; \beta)
\label{Partition_Function}
\end{equation}
at inverse temperature $\beta = 1/(k_{B}T)$. In the above, $\vec{R}=\{ \vec{r}_{1}, \vec{r}_{2}, ...,\vec{r}_{N}\}$ is the position vector of all $N$ particles' coordinates, where $\vec{r}_{i}$ represents the position of the $i$th particle. One of the key quantities present in both Equations \eqref{Observable} and \eqref{Partition_Function} is $\rho(\vec{R}, \vec{R}^{'}; \beta)$, the finite temperature density matrix. In position space, the density matrix may be expanded into

\begin{eqnarray}
\rho(\vec{R}, \vec{R}^{'}; \beta) &=& \langle \vec{R}| e^{-\beta \hat{H}} | \vec{R}^{'} \rangle \nonumber \\ &=&
\sum_{i} \phi_{i}^{*}(\vec{R}) \phi_{i}(\vec{R}^{'}) e^{-\beta E_{i}},
\label{ThreeD_DensityMatrix}
\end{eqnarray}
where $\hat{H}$ is the Hamiltonian that describes the system, and $E_{i}$ and $\phi_{i}$ respectively denote its eigenvalues (energies) and eigenfunctions (wave functions). Although obtaining an exact expression for the density matrix at an arbitrary temperature would require diagonalizing the many body Hamiltonian, a task generally beyond our current computational capabilities, the density matrix may be simplified into more tractable expressions via Trotter factorization. In general, the convolution of two density matrices remains a density matrix such that

\begin{equation}
\rho(\vec{R}_{1}, \vec{R}_{3}; \beta_{1}+\beta_{2}) = \int d\vec{R}_{2}\rho(\vec{R}_{1}, \vec{R}_{2}; \beta_{1}) \rho(\vec{R}_{2}, \vec{R}_{3}; \beta_{2}), 
\label{Convolution_Equation}
\end{equation}
which implies that the expression for the density matrix at inverse temperature, $\beta$, may be expressed as the convolution of $M$ density matrices at inverse temperature $\tau = \beta/M$

\begin{eqnarray}
\rho(\vec{R}_{0}, \vec{R}_{M}; \beta ) &=& \int ... \int d\vec{R}_{1} d\vec{R}_{2}...d\vec{R}_{M-1} \rho(\vec{R}_{0}, \vec{R}_{1}; \tau) \nonumber \\
&&\rho(\vec{R}_{1}, \vec{R}_{2}; \tau)...\rho(\vec{R}_{M-1}, \vec{R}_{M}; \tau).
\end{eqnarray}
The $M$ different $N$-particle position vectors represent the particle positions at $M$ different so-called ``imaginary times.'' Because $\vec{R}_{0}$ is the same as $\vec{R}_{M}$ in the expression for the partition function, according to the quantum-classical isomorphism, the particles may be thought of as polymers constructed of $M$ different beads linked together. If $\tau$ is sufficiently small, assuming the system Hamiltonian may be split into kinetic ($\hat{K}$) and potential pieces ($\hat{V}$), $\hat{H}=\hat{K} + \hat{V}$, the Suzuki-Trotter factorization\cite{Trotter_1959,Kato_1974} may be used to break up the exponential of the Hamiltonian into corresponding kinetic and potential exponentials

\begin{equation}
e^{-\tau \hat{H}} = e^{-\tau (\hat{K} + \hat{V})} \approx e^{-\tau \hat{K}} e^{-\tau \hat{V}} 
\end{equation}
such that the position-space density matrices may also be broken into

\begin{equation}
\rho(\vec{R}_{0}, \vec{R}_{2}; \tau) \equiv \int d\vec{R}_{1} \langle \vec{R}_{0}| e^{-\tau \hat{K}} | \vec{R}_{1} \rangle \langle \vec{R}_{1}|e^{-\tau \hat{V}} | \vec{R}_{2} \rangle.
\end{equation}
Because potential operators are generally diagonal in the position representation, evaluating the position density matrix is trivial

\begin{equation}
\langle \vec{R}_{1}|e^{-\tau \hat{V}} | \vec{R}_{2} \rangle = e^{-\tau \hat{V}(\vec{R}_{1})} \delta(\vec{R}_{2}-\vec{R}_{1}). 
\end{equation}
The evaluation of the kinetic density matrix is less trivial and requires finding the eigenvalues and eigenfunctions of $\hat{K}$. Assuming periodic boundary conditions and the $\alpha=2$, non-fractional form for the kinetic operator

\begin{equation}
\hat{K} = \frac{-\hbar^{2}}{2m} \nabla^{2} = -\lambda \nabla^{2}, 
\label{Kinetic_Operator}
\end{equation}
in which $\Delta$ denotes the Laplacian, $\nabla$ denotes the differential operator, and $\lambda = -\hbar^{2}/(2m)$, the eigenvalues of the three-dimensional kinetic operator are $\lambda \vec{K}_{n}^{2}$ and the eigenfunctions are $L^{-3N/2}e^{i\vec{K}_{n}\vec{R}}$, with $\vec{K}_{n} = 2 \pi \vec{n}/L$. The kinetic density matrix may therefore be expressed as

\begin{eqnarray}
 \langle \vec{R}_{0} | e^{-\tau \hat{K}} | \vec{R}_{1} \rangle  &=&
\sum_{\vec{n}} L^{-3N} e^{-\tau \lambda K_{n}^{2} - i \vec{K}_{n} (\vec{R}_{0}-\vec{R}_{1}) } \nonumber \\ &=&
(4 \pi \lambda \tau)^{-3N/2} e^{\frac{-(\vec{R}_{0}-\vec{R}_{1})^{2}}{4 \lambda \tau}}. 
\label{Usual_Kinetic_Operator}
\end{eqnarray}
As discussed in more detail in Section \ref{sec:Moves}, the finite temperature partition function may then be sampled via Monte Carlo by sampling the Gaussian kinetic density matrix probability distribution function for new particle coordinates and then using the potential density matrices to accept/reject those coordinates according to their potential energy contributions. As a prelude to the subsequent discussion, it should be noted that Equation \eqref{Usual_Kinetic_Operator}, and the eigenvalues and eigenfunctions used to evaluate it, is only valid for the $\nabla^{2}$ Laplacian. If the differential operator is of a different power, the Suzuki-Trotter factorization is still valid (see Appendix II), but new expressions for the density matrix are required. These expressions are derived and discussed below. 

\subsection{\label{sec:FSWE} The Fractional Schr\"odinger Equation}

As first presented by Laskin,\cite{Laskin_1,Laskin_2,Laskin_3,Laskin_4} the fractional Schr\"odinger Equation is a generalization of the Schr\"odinger Equation in which the Hamiltonian is generalized to a fractional form, $\hat{H}_{\alpha}$, that contains a fractional Laplacian, $\Delta^{\alpha/2}$. In one dimension, $\hat{H}_{\alpha}$, may be expressed as

\begin{equation}
\hat{H}_{\alpha} = - D_{\alpha} \hbar^{\alpha} \Delta^{\alpha/2} + \hat{V}(x).
\label{Fractional_Hamiltonian}
\end{equation}
In the above, $\hat{V}(x)$ denotes the potential and $D_{\alpha}=\left(1/2m\right)^{\alpha/2}$ denotes the constant that falls in front of the Laplacian. In general, $D_{\alpha}$ must have SI units of

\begin{equation}
[D_{\alpha}] = \frac{ m^{2-\alpha} kg^{1-\alpha}}{s^{2-\alpha}}. 
\end{equation}
The fractional exponent, $\alpha$, may in principle assume any positive value, including non-integer, fractional values, as its name implies.\cite{Laskin_1,Laskin_2,Laskin_3,Laskin_4} For reference, the fractional Hamiltonian reduces to the conventional case when $\alpha=2$. Please also note that in the next few sections we consider the one-dimensional version of the Hamiltonian to simplify our discussion of the derivation of the fractional density matrix; we will return to a discussion of the full three-dimensional form in later sections.

Most previous work on the fractional Schr\"odinger Equation has explored the properties of fractional Hamiltonians with potentials that are analytically soluble, including the delta potential,\cite{Oliveira_Costa_JMP_2010}, the fractional hydrogen atom,\cite{Laskin_2} the fractional oscillator,\cite{Laskin_2} and the fractional square well.\cite{Guo_Fractional,Dong_Fractional,Secchi_Fractional,Felmer_Fractional} In addition to the fact that these models manifest atypical physics, such as atypical energy spectra, they are also of intrinsic mathematical interest. Considering that that the ordinary Schr\"odinger equation is already extremely challenging to solve, the inclusion of a (nonlocal) fractional derivative makes equation \eqref{E:FSE} even less tractable.
More recently, it has been suggested that fractional Hamiltonians may naturally arise from strong many body interactions, which prompted a mean field exploration of fractional Bose-Einstein condensation.\cite{Kleinert_Paper} Nevertheless, no robust algorithm for examining the properties of the fractional Schr\"odinger Equation for arbitrary potentials has yet been proposed.  

\color{black}
\subsection{\label{sec:Solutions_FSWE} Spectrum of the Fractional Laplacian}
\color{black}

In order to generalize the Path Integral Monte Carlo algorithm discussed in Section \ref{sec:Review_PIMC} to fractional Hamiltonians, expressions for the fractional density matrix must be obtained. As discussed above, this requires finding an analytical form for the kinetic (non-interacting) density matrix that is also amenable to sampling. This can be achieved by finding the eigenvalues and eigenfunctions of the fractional Laplacian. The eigenvalue spectrum of the Riesz fractional Laplacian in one dimension is given by
\begin{eqnarray}
\Delta^{\alpha/2} \cos (Cx) &=&  -|C|^{\alpha} \cos(C x) \\
\Delta^{\alpha/2} \sin (Cx) &=&  -|C|^{\alpha} \sin(C x) 
\end{eqnarray}
for any $C \in \mathbb{R}$ in the unbounded case.\cite{Hermann_Fractional} In fact, this can be viewed as the definition of the real power of the Laplacian operator via the spectral theorem.\cite{Rudin_Functional}

\renewcommand{\thefootnote}{$\ddagger$} 
The standard boundary conditions used in condensed phase simulations are periodic boundary conditions. The introduction of periodic boundary conditions restricts the continuous family of eigenfunctions to a discrete family of eigenfunctions that satisfy the required periodicity.\footnote{We warn the reader that this is not the case for the Riesz fractional Laplacian in a bounded domain with Dirichlet (or Neumann) boundary conditions. There, the eigenfunctions of the Riesz fractional Laplacian are not those of the standard Laplacian in the same domain. See Chapter 12 of Hermann\cite{Hermann_Fractional} for a discussion of this issue, which is due to the nonlocality of fractional derivatives. The direct spectral power of the fractional Laplacian in a bounded domain (that preserves the eigenfunctions of the standard operator) results in a different fractional Laplacian, the \emph{spectral fractional Laplacian}. However, this is not the same fractional Laplacian that is prescribed by the derivation of the fractional Schr\"odinger Equation of Laskin -- except on $\mathbb{R}^n$.} For a periodic box of length $L$, $C=2\pi k/L$, where $k=1,2,3,...$,  the related eigenvalues of the Laplacian are therefore
\begin{equation}
\lambda = - |C|^{\alpha} =  - \left( \frac{ 2 \pi k}{L} \right)^{\alpha} 
\end{equation}
and the related eigenfunctions are $\sin(Cx)$ and $\cos(Cx)$. As in the standard derivation for the particle in a box Hamiltonian,\cite{Krauth_Statistical} the sums and differences of these trigonometric functions may be combined and normalized to yield eigenfunctions of the form 

\begin{equation}
\phi_{k}^{per,L}(x) = \sqrt{ \frac{1}{L}} e^{-iCx}. 
\end{equation}
Inserting these expressions into Equation \eqref{Fractional_Hamiltonian}, the final Hamiltonian eigenvalues (energies) may be obtained

\begin{eqnarray}
 -D_{\alpha} \hbar^{\alpha} \Delta^{\alpha/2} \phi_{k}^{per,L}(x) &=& D_{\alpha}  \hbar^{\alpha} |C|^{\alpha}  \sqrt{ \frac{1}{L}} e^{-i C x} \nonumber \\ &=& E_{k}^{per,L} \sqrt{ \frac{1}{L}} e^{-i C x} . 
\end{eqnarray}
Thus, 

\begin{equation}
E_{k}^{per, L} = D_{\alpha} \hbar^{\alpha} |C|^{\alpha} = D_{\alpha} \left| \frac{ 2 \pi k \hbar }{L} \right|^{\alpha}.
\end{equation}
Consequently, the energies in the fractional periodic setting are similar in form to the $\alpha=2$, non-fractional case, except that the frequencies are now raised to a fractional power instead of being squared. 

\subsection{\label{sec:Fractional_Density_Matrix} Derivation of the Fractional Density Matrix}

Following Ceperley and Krauth,\cite{Ceperley_Rev_Mod_Phys,Krauth_Statistical} the kinetic density matrix in one dimension may be obtained along the same lines as in Equation \eqref{Usual_Kinetic_Operator} by summing over the Hamiltonian's eigenvalues and eigenfunctions

\begin{eqnarray}
\rho^{per, L}(x,x',\beta)  &=& \sum_{k=-\infty}^{\infty} \phi_{k}^{per,L}(x) e^{-\beta E_{k}^{per,L}} \phi_{k}^{*,per,L}(x') \nonumber \\
&=& \frac{1}{L} \sum_{k=-\infty}^{\infty} e^{iC(x'-x)} e^{-\beta D_{\alpha} \hbar^{\alpha} |C|^{\alpha}} 
\label{Initial_Fractional_Density_Matrix}
\end{eqnarray}
As usual, these sums may be turned into integrals by substituting the implied $\Delta k$ in the summation with $\Delta k = \Delta C L/(2 \pi)$ and letting the box size, $L$, go to infinity
\begin{multline}
\label{integral}
\rho^{per, L}(x,x',\beta) = \\
\lim_{L\rightarrow \infty} \frac{1}{L} \sum_{C=...,-\frac{2\pi}{L},0,\frac{2\pi}{L},...} \left( \frac{\Delta C L}{2 \pi} \right) e^{i C  (x-x') } e^{-\beta D_{\alpha} \hbar^{\alpha} |C|^{\alpha}} \\ 
= \frac{1}{2 \pi} \int_{-\infty}^{\infty} dC e^{i C  (x-x') } e^{-\beta D_{\alpha}  \hbar^{\alpha} |C|^{\alpha}}. 
\end{multline}
If $\alpha=2$, this integral boils down to the usual Gaussian case in which completing the square yields the final line of Equation \eqref{Usual_Kinetic_Operator}. 

\color{black}For more general values of $\alpha$, it is impossible to compute this integral in closed form, but it can be represented as a special function defined by a power series. \color{black}
The integral may be identified as the characteristic function of the generalized normal random variable $X_{GN}$. This is the random variable given by the probability density function
\begin{equation}\label{generalized_normal}
f_{\kappa}(\mu, \sigma; x) = \xi(\kappa) = \frac{ \kappa}{2 \sigma \Gamma(1/\kappa)} e^{-\left| \frac{x-\mu}{\sigma} \right|^{\kappa} }. 
\end{equation}
The characteristic function of $\xi(\kappa)$ for $\kappa>1$ 
was first computed explicitly in terms of special functions by Pogany and Nadarajah in 2006.\cite{Pogany_2010} It may be expressed as
\begin{equation}
E\{ e^{itX_{GN}} \} = 
\frac{ \sqrt{\pi} e^{it\mu} }{\Gamma(1/\kappa)} \Psi^{1}_{1}\left[ (1/\kappa, 2/\kappa); (1/2, 1); -\frac{(\sigma t)^{2} }{4} \right],
\label{characteristic}
\end{equation}
where $\Psi^{1}_{1}$ denotes the Fox-Wright generalized hypergeometric function. This Fox-Wright function is a  generalization of the generalized hypergeometric function, and is a special case of the Fox $H$-function. It is specified by an arbitrary number, ${p+q}$, of parameters, and is 
defined as the series
\begin{multline}\label{fox-wright}
\Psi^{p}_{q} \left[ (\alpha_{1}, A_{1}), ..., (\alpha_{p}, A_{p}); (\beta_{1}, B_{1}), ..., (\beta_{q}, B_{q}); z \right] \\
= \sum_{n=0}^{\infty} \frac{ \prod_{j=1}^{p} \Gamma(\alpha_{j} + A_{j}n) }{ \prod_{j=1}^{q} \Gamma(\beta_{j} + B_{j}n) } \frac{z^{n}}{n!}. 
\end{multline}
Comparing the generalized normal distribution in the expression for the density matrix given by Equation \eqref{integral} with the definition of the generalized normal distribution given by Equation \eqref{generalized_normal},  we have $|x-\mu| = |C|$, $\sigma = \frac{1}{\beta^{1/\alpha} D_{\alpha}^{1/\alpha} \hbar}$, and $\mu = 0$. Making these substitutions, we obtain
\color{black}
\begin{eqnarray}
&{\rho^{per, L}}&(x,x',\beta) \nonumber\\ 
&=& \frac{1}{2 \pi} \int_{-\infty}^{\infty} dC e^{i C  (x-x') } e^{-\beta D_{\alpha}  \hbar^{\alpha} |C|^{\alpha}} \nonumber \\ 
&=& \frac{1}{2\pi} \frac{ 2 \Gamma(1/\alpha)}{ \beta^{1/\alpha} D_{\alpha}^{1/\alpha} \hbar \alpha} \frac{ \sqrt{\pi} e^{i (x-x') \mu}}{\Gamma(1/\alpha)} \nonumber \\
&& \quad \times \Psi^{1}_{1} \left[ (1/\alpha, 2/\alpha); (1/2, 1); \frac{ - (x-x')^{2}}{4 \beta^{2/\alpha} D_{\alpha}^{2/\alpha} \hbar^{2}}  \right] \nonumber \\
&=&  \frac{ 1 }{ \sqrt{\pi} \beta^{1/\alpha} D_{\sigma}^{1/\alpha} \hbar \alpha} \nonumber \\
&& \quad  \times \sum_{n=0}^{\infty} \frac{ \Gamma(1/\alpha + (2/\alpha) \,n) }{ \Gamma(1/2 + n)} \frac{ \left[  \frac{ - (x-x')^{2}}{4 \beta^{2/\alpha} D_{\alpha}^{2/\alpha} \hbar^{2}} \right]^{n} }{n!}. 
\label{distribution}
\end{eqnarray}
\color{black}
The generalized expression for the fractional kinetic density matrix is therefore proportional to the Fox-Wright  function of argument $(x-x')^2$. \color{black} Similar representations were obtained by Laskin\cite{Laskin_4} for the free-particle density matrix on $\mathbb{R}^n$ without boundary conditions.  Laskin used the more general {Fox $H$-function} in his representation, which for the given indices reduces to the {Fox-Wright} function shown here. In another context, similar results were derived by Mainardi and Pagnini,\cite{Mainardi_Pagnini} where the integral form of \eqref{distribution} arises as the fundamental solution to the space-fractional diffusion equation. 

At this point, an efficient way to sample this Fox-Wright density is required to properly sample the kinetic density matrix. \color{black}

\subsection{\label{sec:Computing_Density_Matrix}Computing and Sampling the Fractional Density Matrix}

The Fox-Wright density given by Equation  \eqref{distribution} is interpreted as a probability
density function (PDF) of the jump distance, $|{x-x'}|$, for the random walkers. 
For $\alpha = 2$, the Fox-Wright density reduces to a
Gaussian, which is typically sampled using the Box-Muller transformation.\cite{BoxMuller} 

\begin{figure}[h]
  \centering
  \textbf{Discrete Inverse Transform Sampling}
    \includegraphics[width=\columnwidth]{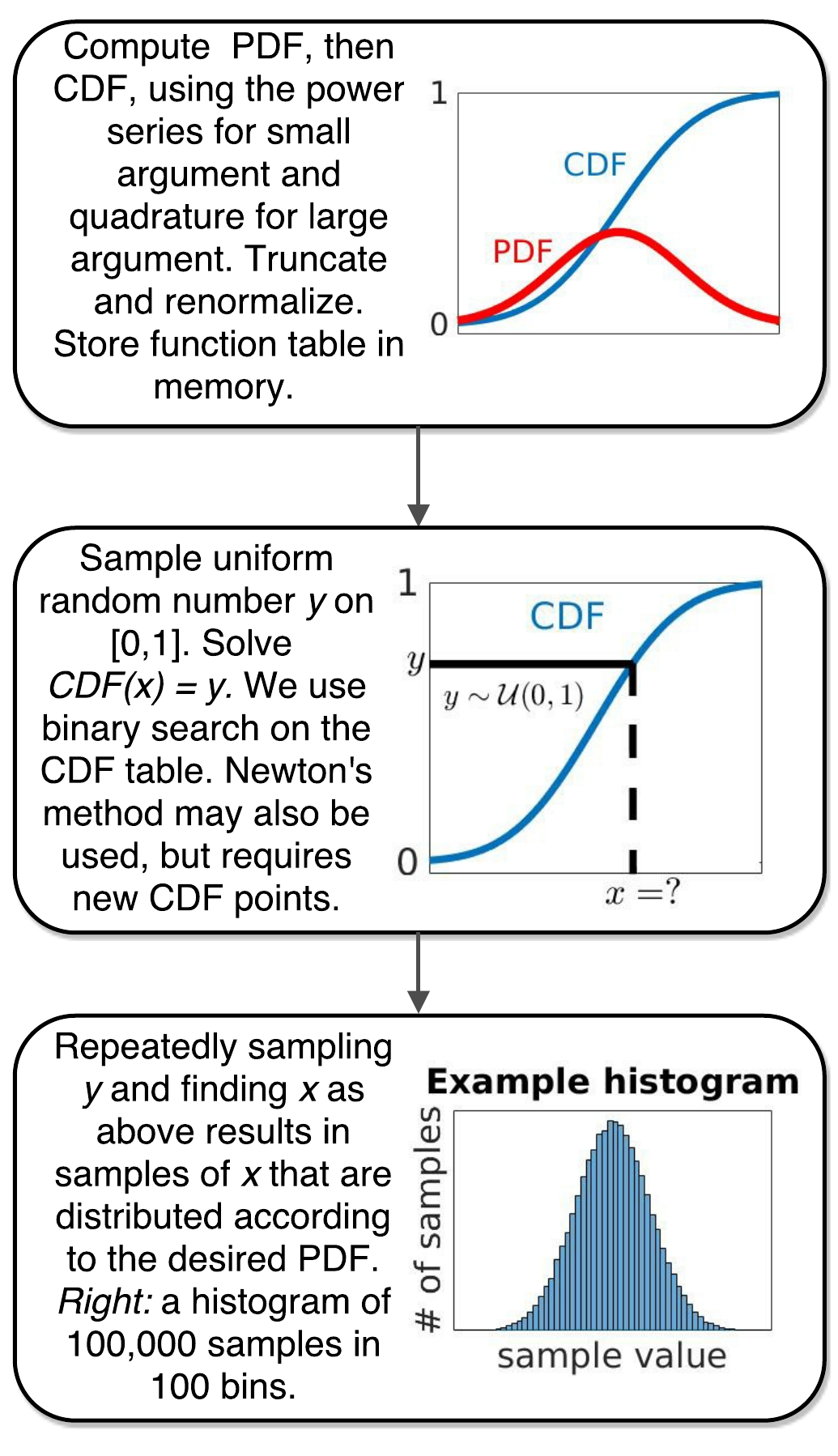}
\caption{A summary of the discrete inverse-transform sampling method used in this work to sample free particle density matrices.}
\label{flowchart}

\end{figure}

\color{black}
For simplicity and portability, we have used discrete inverse transform sampling to sample the distribution \eqref{distribution}. The method is entirely sufficient for the initial computations presented here. It is both fast and the most versatile method; 
since it does not rely on any special transformations, it can be used to sample any random variable for which the probability density function can be computed reliably. 
\color{black}
Our method is summarized in Figure \ref{flowchart} and described in full detail in this section. 

The parameters 
$\beta$, $D_\alpha$, and $\hbar$ are not essential to the sampling methodology. We begin by reducing them to unity. If the density given in Equation \eqref{integral} is denoted by ${\rho_{\beta, D_\alpha, \hbar}}$, then
\begin{equation}\label{rho_1_1_1}
\rho_{1,1,1}(x-x') = 
\frac{1}{2\pi} \int_{-\infty}^{\infty}
e^{iC(x-x')} e^{-|C|^\alpha} dC. 
\end{equation} 
By rescaling, ${\rho_{\beta, D_\alpha, \hbar}}$ may be reduced to $\rho_{1,1,1}(x-x')$. 
Letting  
${C = C'(\hbar^{\alpha} \beta D_\alpha)^{-1/\alpha}}$
and
${dC = dC' (\hbar^{\alpha} \beta D_\alpha)^{-1/\alpha}}$,
so that
$\beta D_\alpha \hbar^{\alpha} |C|^\alpha = |C'|^{\alpha}$, 
we get  
\begin{eqnarray}
&& \rho_{\beta, D_\alpha, \hbar} (x-x') \\ 
&=& \frac{1}{2\pi} \int_{-\infty}^{\infty}
(\beta D_\alpha \hbar^\alpha)^{-1/\alpha} 
e^{-C' (\beta D_\alpha \hbar^\alpha)^{-1/\alpha}(x-x')  }
e^{-|C'|^{\alpha}}
dC' \nonumber \\
&=& \frac{1}{2\pi} \int_{-\infty}^{\infty}
(\beta D_\alpha \hbar^{\alpha})^{-1/\alpha}
e^{-C' \left( \frac{x-x'}{(\beta D_\alpha \hbar^\alpha )^{1/\alpha}} \right)  }
e^{-|C'|^{\alpha}}
dC' \nonumber \\
&=&
(\beta D_\alpha \hbar^\alpha)^{-1/\alpha} 
\rho_{1,1,1}
\left(
\frac{x-x'}{(\beta D_\alpha \hbar^\alpha)^{1/\alpha}} 
\right) \nonumber
\end{eqnarray}
Therefore, in what follows, we can assume without 
loss of generality that  ${\beta = D_\alpha = \hbar = 1}$. 

As mentioned, the PDF ${\rho_{1,1,1}}$ can be sampled using inverse transform sampling, which requires that the related cumulative density function (CDF) be inverted. The general result is that, if a random variable $X$ has CDF $F$, then $X = F^{-1}[\mathcal{U}(0,1)]$ in distribution,\cite{Rubinstein_Monte_Carlo} where $\mathcal{U}(0,1)$ denotes the uniform random variable on $(0,1)$. In our case, the CDF of the random variable given by $\rho_{1,1,1}$ does not admit analytic inversion, so various numerical methods can be used. Options include binary search on a finely-discretized function table or a Newton method. Since PIMC simulations require samples of $x-x'$ each of the millions of times a bead is displaced, the numerical inversion must be carried out as efficiently as possible. 
Any method for inverting the CDF will require accurate computation of the CDF, and therefore the PDF ${ \rho_{1,1,1} }$. 

There are two direct ways to compute the PDF of $\rho_{1,1,1}$:
direct quadrature of the integral in the first line of Equation \eqref{distribution}
or truncation of the Taylor series in the last line of Equation \eqref{distribution}.
The truncated Taylor series is a convenient method when the argument of the distribution is small. For larger arguments, the series (which is an expansion
at $x = 0$) requires increasingly higher-degree terms to be accurate, but this leads quickly to numerical blow-up.  
The combination of high-degree monomials,
large ${x}$, and gamma functions and factorials with large arguments 
results in overflow and precision issues. One work around is 
the following numerical ``order of operations'' trick:
\begin{equation}
\frac{x^n}{n!} = 
\left( \frac{x}{1} \right)
\left( \frac{x}{2} \right)
...
\left( \frac{x}{n-1} \right)
\left( \frac{x}{n} \right).
\end{equation}
This prevents the overflow arising from computing the very large numbers
${x^n}$ and ${n!}$ separately before dividing by pairing
each large factor ${x}$ with a large factor ${k}$, $k = 1, 2, ... n$.
With this trick, one can use the Taylor series to compute the Fox-Wright function
${\rho_{1,1,1}}$ accurately and quickly for arguments up to about 10, 
using 100 terms in the series. \color{black}Beyond that point, numerical issues become
more difficult to resolve and the computational cost  too high. \color{black}

Thus, for larger arguments, direct quadrature becomes more favorable. 
Due to the rapid decay of ${e^{-|C|^\alpha}}$, the truncated integral
\begin{multline}\label{trapezoid}
\frac{1}{2\pi} \int_{-L}^L e^{iC(x-x')}e^{-|C|^\alpha} dC \\
=
\frac{1}{2\pi} \int_{-L}^L \cos{\left[C(x-x') \right]}e^{-|C|^\alpha} dC
\end{multline}
converges rapidly in ${L}$. At ${L \sim 50}$, the integrand is far below machine precision. Note that in Equation \eqref{trapezoid}, the complex exponential has been replaced by a cosine because the related sine term is odd and therefore does not contribute to the final integral. The truncated integral can then
be approximated by any quadrature rule, e.g., the 
trapezoid rule. 
For large $x-x'$, this is appealing as there are no overflow/precision 
problems. The only hazard is that the \color{black}frequency of the integrand \color{black} is proportional
to ${x-x'}$, so the quadrature must be performed with care to ensure that the result is converged in the number of quadrature points for large arguments. Quadrature is relatively costly in terms of computational time.

Our final algorithm for sampling is as follows. Treating the PDF ${\rho_{1,1,1}}$ as discrete, we compute the CDF on ${[-100,100]}$ in intervals of 0.01. This is done using the power series for ${|x| < 2}$ and quadrature for ${2 < |x| < 100}$. The choice of at which value to transition from using the power series to using quadrature is immaterial as long as both techniques are accurate in the given region. For ${|x| > 100}$, where the Fox-Wright density approaches machine precision, we truncate. The resulting CDF is normalized to correct the minute mass lost by truncation and stored in memory. To generate a random sample $x$, we use binary search on the stored table to invert the equation $y = \text{CDF}(x)$, where $y$ is drawn from a uniform distribution on ${[0,1]}$. This method incurs high overhead when initializing the discrete Fox-Wright CDF on a fine grid, but avoids computation of the Fox-Wright density $\rho_{1,1,1}$ after initialization. Because the Fox-Wright CDF is computed before Monte Carlo moves are made, this approach avoids the computational bottlenecks the computation of the CDF would otherwise present.

It is not necessary to fully discretize the density matrix; in principle, one could use a Newton solver with a coarse binary search for the initial guess to invert the CDF. However, we find the computational overhead to be similar to that needed for discretization/a binary search. In the fully discrete approach, after initialization, no expensive computations of the Fox-Wright density are ever performed, while in the Newton method, the Fox-Wright density must be computed several times at new values to generate each sample. If a specific computation proves to be extremely sensitive to the heavy tails of the Fox-Wright distribution, and it is desired to avoid truncation, the more expensive Newton method may be required.

\color{black}
We mention that the inversion of the Fox-Wright CDF could be simplified and accelerated if a simple asymptotic expression as a power law were known. Since this function enters into a power-law regime fairly quickly for ${\alpha \neq 2}$ (see Figures \ref{First_Distribution_Figure} and \ref{Second_Distribution_Figure}), this is a reasonable hope. Any asymptotic expansion that admits closed-form, analytic inversion would make inversion of the Fox-Wright CDF trivial for many samples. Although there exists a vast literature on asymptotic expansions for such special functions, we have not found the desired expression that will serve this purpose.
\color{black}

\color{black}
Finally, we remark that inverse-transform sampling is not the only method that applies here. The CMS (Chalmers-Mallows-Stuck) transform method \cite{Chalmers_JASA} for sampling $\alpha$-stable variables does not rely on the explicit 
computation of the PDF given by Equation \eqref{distribution}. The CMS method can be thought of the closest approach to the Box-Muller transform for stable procsses. However, as the developers of that method note,\cite{Chalmers_JASA} it is not clear what method is best for very large-scale Monte Carlo simulations. This is an interesting point for future survey. Regardless, as shown below, sample generation using discrete inverse transform sampling is quite fast after the CDF has been initialized, generating $\mathcal{O}(10^4)$ samples per second on a single core. Accurate tables of the CDF can be saved, so initialization is not required for future calculations. Moreover since the discrete-inverse transform method does not rely on special transforms, it can be readily applied to an arbitrary distribution. This is the main advantage of inverse-transform sampling. For example, if it was desired to use tempered distributions\cite{Meerschaert_JCOMP} (see section \ref{sec:Conclusions}) rather than pure $\alpha$-stable distributions, it would only be required to swap the PDF subroutines. Thus, the implementation of the inverse-transform method is worthwhile for any future work on fractional PIMC.
\color{black}

\subsection{\label{sec:Performing_PIMC} Path Integral Monte Carlo Simulations Based Upon the Fractional Density Matrix}

\subsubsection{\label{sec:Moves} Performing PIMC Moves}

As described in greater detail elsewhere,\cite{Boninsegni_Permutation} in Path Integral Monte Carlo, a random walk is performed through the space of $N$-particle paths at the $M$ different time slices, $\{\vec{R}_{1}, \vec{R}_{2},...,\vec{R}_{M}\}$. A random walk may be constructed by randomly displacing the positions of the different beads within the particles. While many different move possibilities exist,\cite{} one of the simplest moves is to displace a subset of the beads in the $M$-bead path of a given particle. Let the portion of a particle $i$'s path to be updated be denoted by $\vec{r}_{i,k+1}...\vec{r}_{i,k+s-1}$, where $s$ is the number of beads in the path to be displaced. Let $\vec{X}=\{\vec{R}_{0},\vec{R}_{1},...,\vec{R}_{M}\}$ denote the old set of particle bead coordinates and let $\vec{X}^{'}=\{\vec{R}_{0},...\vec{R}_{k},\vec{R}_{k+1}^{'}...\vec{R}_{k+s-1}^{'},\vec{R}_{k+s},...,\vec{R}_{M}\}$ denote the updated (new) set of particle bead coordinates. Then, based upon Equation \eqref{Partition_Function}, the probability, $P$, that this displacement will be accepted from sampling the partition function is

\begin{eqnarray}
P &=& \frac{ \prod_{j=0}^{s-1} \rho_{K}(\vec{r}_{k+j}', \vec{r}_{k+j+1}', \Delta \tau)  }{ \prod_{j=0}^{s-1} \rho_{K}(\vec{r}_{k+j}, \vec{r}_{k+j+1}, \Delta \tau) } \nonumber \\ && \quad \times \frac{ e^{-\Delta \tau \sum_{j=1}^{s-1} V(\vec{R}_{k+j}') }}{ e^{-\Delta \tau \sum_{j=1}^{s-1} V(\vec{R}_{k+j}) } } 
\frac{ T(\vec{X}' \rightarrow \vec{X})}{ T(\vec{X} \rightarrow \vec{X}^{'})}. 
\label{Move_Prob}
\end{eqnarray}
Here, $T(\vec{X}^{'} \rightarrow \vec{X})$ represents the transition probability, which can be selected with great freedom. The most convenient choice for the transition probability is to set it equal to the product of the kinetic density matrices, so that it can be evaluated using the expressions derived above and can cancel out other contributions to $P$. If this choice is made, 

\begin{equation}
T(\vec{X} \rightarrow \vec{X}^{'}) = \prod_{j=0}^{s-1} \rho_{K}(\vec{r}_{k+j}', \vec{r}_{k+j+1}', \Delta \tau),
\end{equation}
and Equation \eqref{Move_Prob} reduces to

\begin{equation}
P=e^{-\Delta \tau \sum_{j=1}^{s-1} (V(\vec{R}_{k+j}^{'})-V(\vec{R}_{k+j}))}.
\end{equation}
These equations imply that, in a practical PIMC simulation, coordinates should first be sampled from the product of the kinetic density matrices (the length of that product depends upon the number of beads to be moved) and then the move should be accepted/rejected based upon the ratio of the potential energy in the new configuration to the potential energy in the old configuration. 
There are many ways to sample the transition probability and construct multiple bead moves.\cite{Sprik_Staging,Ceperley_Rev_Mod_Phys,Doll_Fourier_1} One of the most common ways to propose multi-bead moves in the typical, $\alpha=2$ case is to use the staging algorithm,\cite{Sprik_Staging} which samples a hierarchy of position moves based upon the fact that a product of Gaussians is a Gaussian. Because our fractional kinetic density matrices are not in general Gaussians and products of Fox-Wright functions do not yield Fox-Wright functions, the staging algorithm cannot be straightforwardly adapted to this formalism. For the purpose of this paper (we discuss alternatives capable of more efficiently sampling multi-bead and permutation moves may in our Conclusions), we therefore use single-bead and center of mass moves, despite the ergodicity problems single-bead moves may cause at low temperatures, particularly for strong interactions. For center of mass moves, all beads in each particle are moved at once, as usual. For single-bead moves, a bead on a particle is randomly selected. New unscaled $x$', $y$', and $z$' coordinates are then independently sampled from their Fox-Wright Distributions as described in Section \ref{sec:Fractional_Density_Matrix} and rescaled by 
\begin{equation}
\sigma^{\alpha} = \frac{ \hbar \tau^{1/\alpha}}{2^{1/\alpha} m^{1/\alpha}} 
\end{equation}
to yield coordinates of the correct dimensions. As further detailed in Appendix I, the $x$, $y$, and $z$ coordinates may be independently sampled because the many particle density matrices corresponding to the free fractional Hamiltonian may be re-expressed as the product of single particle density matrices in each dimension. The resulting potential energy from this move is lastly determined and used to accept/reject the proposed move. 

\subsubsection{\label{sec:Observables} Computing Observables}

The fundamental purpose of performing PIMC simulations in most cases is to obtain measures of  different systems' finite temperature observables, such as their energies, radial distribution functions, and radii of gyration. Because such observables as the potential energy and radial distribution functions are diagonal in the position representation and therefore do not directly depend upon the fractional exponent, they can be measured as in the $\alpha=2$ case. Appropriate care must simply be taken for $\alpha<2$ cases because these averages are computed from distributions with heavy tails. 

Computing quantities based upon derivatives of the partition function, such as the kinetic energy, however, is much less straightforward. As derived in Appendix III, expressions dependent on derivatives of the partition function are proportional to slightly more complex Fox-Wright distributions. The kinetic energy, for example, may be written as 
\begin{eqnarray}
&& \left \langle \hat{KE} \right \rangle_{link} = \left \langle \frac{3N}{\tau \alpha} + \frac{2 N}{ \tau \alpha} \left[-C_{\alpha}  (\vec{R}-\vec{R'})^{2}  \right] \nonumber  \right. \\ && 
\left. \left[ \sum_{n=0}^{\infty} \frac{\Gamma(1/\alpha + (2/\alpha) (n+1))}{\Gamma(1/2+(n+1))}  \left[ -C_{\alpha}  (\vec{R}-\vec{R'})^{2}  \right]^{n}/n! \right]/ \right. \nonumber \\
&& \left. \left[ \sum_{n=0}^{\infty} \frac{\Gamma(1/\alpha + (2/\alpha) n)}{\Gamma(1/2+n)}  \left[ -C_{\alpha} (\vec{R}-\vec{R'})^{2}  \right]^{n}/n! \right] \right \rangle_{link} 
\end{eqnarray}
where $C_{\alpha}=\frac{2^{2/\alpha}m^{2/\alpha}}{4 \tau^{2/\alpha} \hbar^{2}}$ and $\langle \rangle_{link}$ denotes the average over links between beads. Note that the additional complexity results from the appearance of factors of $n+1$ as opposed to $n$ in the gamma functions. As a result of this additional complexity, the series is less manageable. It is therefore more convenient to evaluate the thermodynamic estimator of the average kinetic energy based upon the formula \color{black} $\langle KE \rangle = \frac{m}{\beta} \frac{ \partial \ln Z}{\partial m}$ \color{black} and to use a three-point forward difference formula in $m$ to estimate the derivative (see Section \ref{Appendix_1}).      

\section{\label{sec:Results} Results and Discussion}

\subsection{\label{sec:Fractional_Density_Distributions} Fractional Density Matrix Distributions}

\begin{figure}[h]
  \centering
    \includegraphics[width=\columnwidth]{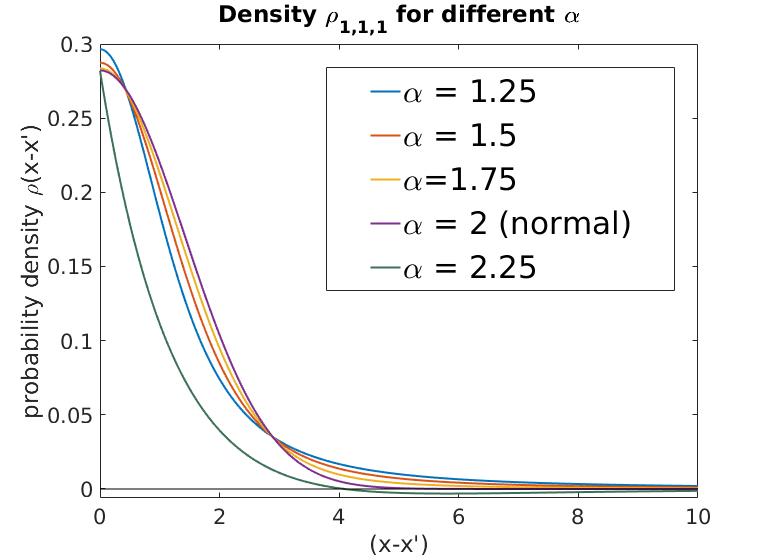}
\caption{Illustration of the fractional density matrix distribution as a function of the average unidimensional interbead distance, $|x-x'|$, for different $\alpha$.}
\label{First_Distribution_Figure}
\end{figure}

\begin{figure}[h]
  \centering
    \includegraphics[width=\columnwidth]{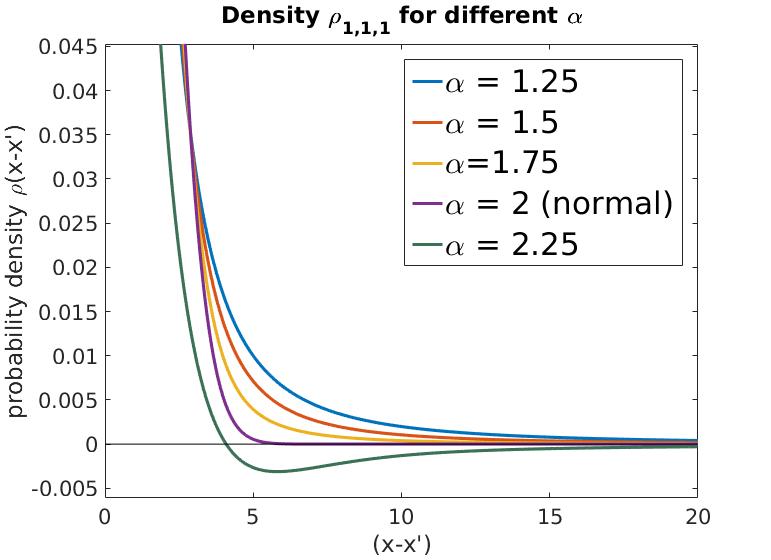}
  \caption{Illustration of the tail behavior of the fractional density matrix. Heavy tails are visible for fractional ${\alpha}$. The case ${\alpha = 2.25}$, which becomes negative, clearly cannot be a probability density function. This issue occurs for all ${\alpha > 2}$, in which case ${\rho}$ has no meaning as a probability density function.}
\label{Second_Distribution_Figure}
\end{figure}

In this section, we compute the Fox-Wright density as described above to analyze its functional form and the performance of our sampling technique. In Figure \ref{First_Distribution_Figure}, we illustrate the overall behavior of the probability density function that represents the uni-dimensional fractional density matrix. The density exhibits ``heavy'' algebraic tails for fractional order 
$\alpha<2$. Compared to the normal case, the fractional cases exhibit a larger probability of both small and large jumps, and a lower probability for medium jumps. In Figure \ref{Second_Distribution_Figure}, we look more closely at the tail behavior. As $\alpha$ decreases, the related probability density functions exhibit heavier and heavier tails. 

In Figures \ref{First_Distribution_Figure} and \ref{Second_Distribution_Figure}, we have also plotted the density ${\rho_{1,1,1}}$ for ${\alpha = 2.25}$ as a curiosity. Of course, ${\alpha}$-stable distributions are only defined for ${0 < \alpha \leq 2}$; this illustration shows that for ${\alpha > 2}$, ${\rho_{1,1,1}}$ takes negative values and is therefore not a density at all. In particular, while ${\alpha < 2}$ corresponds to superduffision, ${\alpha > 2}$ does not correspond in any way to subdiffusion.  This may be surprising at first, but it has the following heuristic explanation. Unlike superdiffusion, one cannot obtain subdiffusion simply by modifying the spatial jump density of Brownian motion. In superdiffusion, the normal jumps are replaced by ``faster'', infinite variance jumps. If one attempts to replace the jumps of Brownian motion by ``slower jumps,'' such a process would necessarily have finite variance and thus, by the central limit theorem, reduce to Brownian motion. Therefore, it is not surprising that for ${\alpha > 2}$, ${\rho_{1,1,1}}$ is meaningless as a density. The only way to obtain subdiffusion is to introduce \emph{waiting times} into the process, which results in a time fractional derivative.\cite{Meerschaert_Stochastic} 

\begin{figure}[h]
  \centering
    \includegraphics[width=\columnwidth]{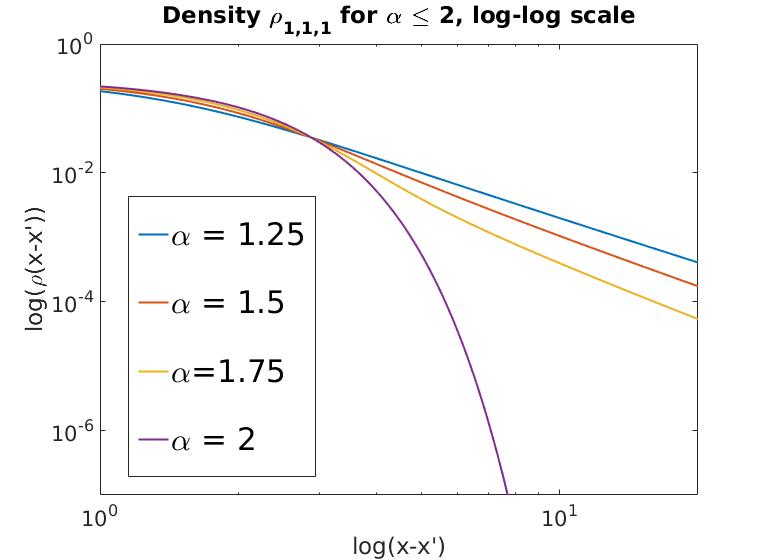}
\caption{When plotted on a log-log scale, it becomes clear that ${\rho_{1,1,1}}$ quickly enters into a power-law regime for fractional ${\alpha}$.}
\label{LogScaleDistribution}
\end{figure}

In Figure \ref{LogScaleDistribution}, we plot the same densities on a log-log scale to better understand their tail behavior. We can see from this that the density enters a power-law regime  fairly early, for $|x-x'| \sim 10$.

\begin{figure}[h]
  \centering
    \includegraphics[width=\columnwidth]{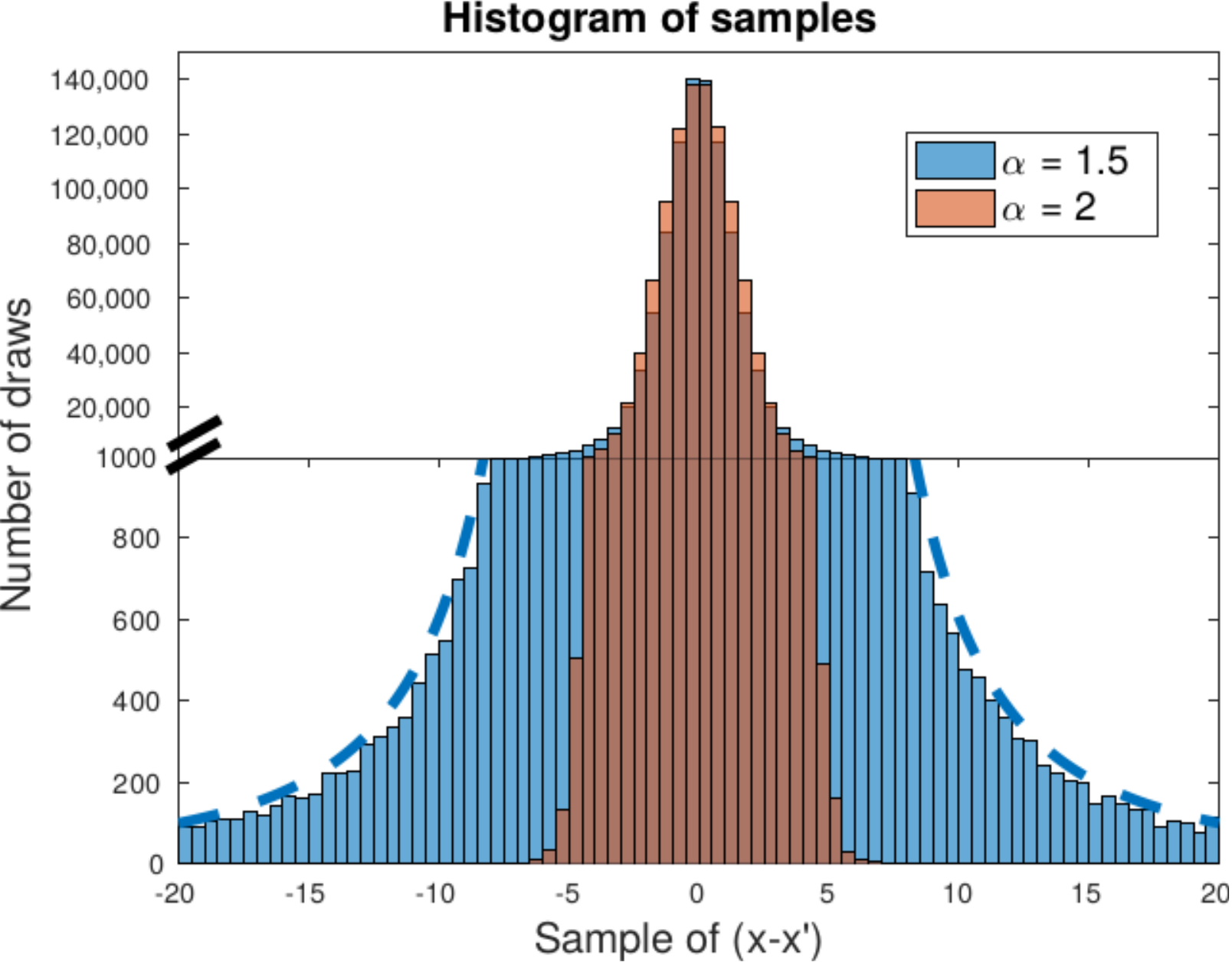}
      \caption{Histograms of samples drawn from the $\alpha=2$ and $\alpha = 1.5$ density matrices using the inverse-transform method described here. In the lower half, the exact PDF for $\alpha = 1.5$ has been plotted as the dashed line to check for correct tail behavior. The density was discretized at $10^4$ points on $[-100,100]$ and $10^6$ samples were drawn. These were sorted into 80 bins on $[-20,20]$. }
\label{Final_Histogram}
\end{figure}

Finally, in Figure \ref{Final_Histogram}, we illustrate the accuracy of the discrete inverse transform sampling method described in Section \ref{sec:Computing_Density_Matrix}. This figure compares histograms representative of the probability density functions obtained for $\alpha=1.5$ and $\alpha=2$ obtained using our inversion algorithm. Note that our inversion method preserves the heavy tails that are central to these distributions.

The numerical costs of our inversion process are presented in Table \ref{T:time}. We see how expensive it is to evaluate the density $\rho$: it takes about 1s to evaluate the density at 25 different points. On the other hand, for the grid used, it takes binary search 1s to generate 6000 samples. Moreover, since binary search is logarithmic in the number of discretization points, similar performance may be expected even for much finer grids. Therefore, if the CDF can be initialized to the desired accuracy, binary search is superior to Newton methods when a very large number of samples is required, i.e., orders of magnitude more than the number of discretization points. A Newton method would require several additional evaluations of $\rho$ for each sample; on a single core, this would generate $<10$ samples per second, compared to the roughly thousands of samples per second for binary search. 

\begin{table}[h]
\centering
\begin{tabular}{|l|l|}
\hline
\textbf{Task}                                                                       & \textbf{\begin{tabular}[c]{@{}l@{}}Time (4.5 Ghz i7-6700K)\\ using one core/thread\end{tabular}} \\ \hline
\begin{tabular}[c]{@{}l@{}}Intialization: $10^4$ \\ discretization pts\end{tabular} & 390 s                                                                                             \\ \hline
\begin{tabular}[c]{@{}l@{}}Binary search sampling: \\ $10^6$ samples\end{tabular}    & 60 s                                                                                              \\ \hline
\textbf{Total}                                                                      & 450 s                                                                                             \\ \hline
\end{tabular}
\caption{Discrete Inverse Transform sample generation run times for $10^6$ samples using $10^4$ discretization points. The probability density function was truncated (set to 0) for $|x| > 100$.}
\label{T:time}
\end{table}

\subsection{\label{sec:Fractional_Kinetic_Energy} Fractional Kinetic Energy}

\begin{figure}[h]
  \centering
   \includegraphics[width=\columnwidth]{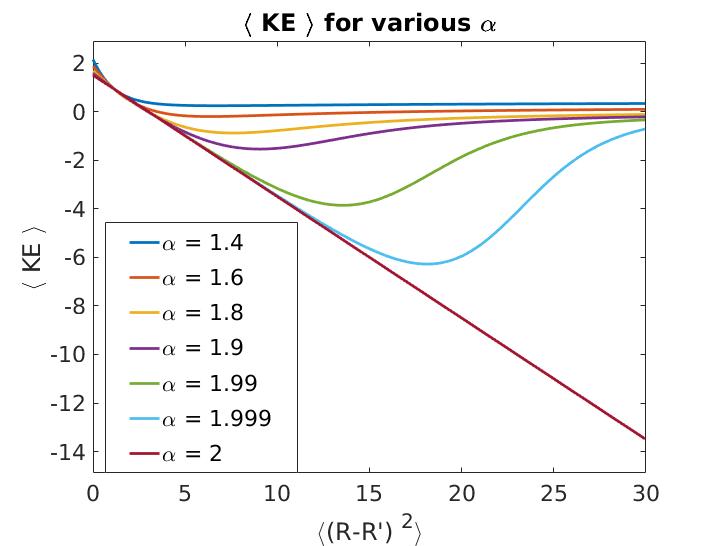}
     \caption{Comparison of the kinetic energy $\langle KE \rangle$ as a function of the interbead distance squared for different $\alpha$. The parameters $\tau, \hbar, N, N_p, m,$ and $\beta$ are all set to $1$ in these figures. }
\label{KE_Curves_1}
\end{figure}

\begin{figure}[h]
  \centering
   \includegraphics[width=\columnwidth]{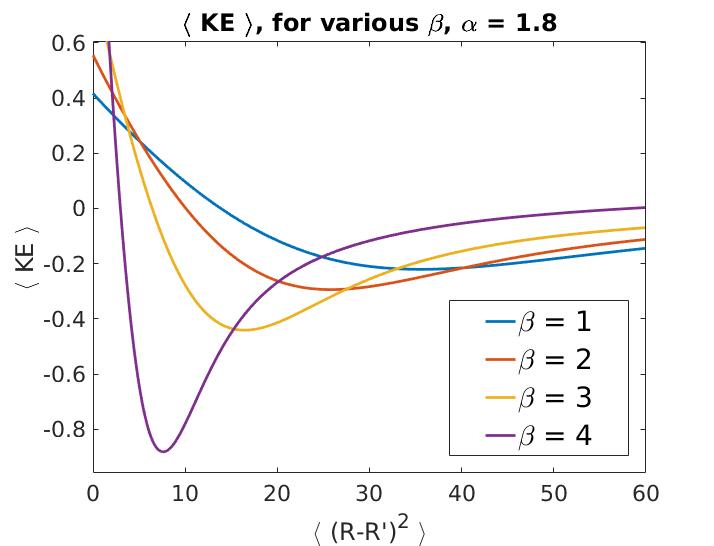}
    \caption{Comparison of the kinetic energy $\langle KE \rangle$ vs interbead distance squared for different $\beta$, using a fixed ${\alpha = 1.8}$. The remaining parameters $\hbar, N, N_p,$ and $m$ are all set to $1$. The parameter $\tau$ satisfies $\tau N_p = \beta$, meaning that it varies with $\beta$.}
\label{KE_Curves_2}
\end{figure}

Using the forward difference scheme described in Appendix III, we analyze the functional form of the kinetic energy as a function of the interbead distance squared for a range of $\alpha$'s. Interestingly, whereas the kinetic energy decreases linearly with interbead distance for $\alpha=2$, it first decreases, then increases, and finally plateaus for smaller $\alpha$, as depicted in Figure \ref{KE_Curves_1}. Similar functional forms are observed for varying $\beta$ in Figure \ref{KE_Curves_2}.   

The plateaus observed in the kinetic energy may be understood by considering the power-law behavior of our Fox-Wright distribution for large argument. In general, the density matrix connecting two consecutive imaginary times, $\rho_{i}(\tau)$ may be approximated as (see Appendix III)

\begin{eqnarray}
&&\rho_{i}(\tau) = \left( \frac{ 2^{1/\alpha} m^{1/\alpha}}{\sqrt{\pi}\tau^{1/\alpha}\hbar \alpha} \right)^{3 N } \times\\
&& \left[ \sum_{n=0}^{\infty} \frac{\Gamma(1/\alpha + 2/\alpha n)}{\Gamma(1/2+n)} \left[ \frac{- ( \vec{R}_{i}-\vec{R}_{i+1})^{2} 2^{2/\alpha} m^{2/\alpha}}{4 \tau^{2/\alpha} \hbar^{2}}      \right]^{n}/n! \right]. \nonumber
\label{KE_Approx}
\end{eqnarray}
Setting all parameters except ${m}$, ${\alpha}$, and $(\vec{R}_{i}-\vec{R}_{i+1})^2$
equal to 1 in the above, we can write
\begin{equation}
\rho_{i}(\tau) \sim (m^{\frac{1}{\alpha}})^{3N} F_\alpha  \left(m^{\frac{1}{\alpha}}   \sqrt{ (\vec{R}_{i}-\vec{R}_{i+1})^2} \right),
\end{equation}
where ${F_\alpha}$ is the Fox-Wright distribution given by Equation \eqref{fox-wright}. We note that Equation \eqref{fox-wright} has its argument 
squared in the Taylor series (see Equation \eqref{characteristic}), so to write $\rho$ in terms of ${F_\alpha}$ we must take the square-root of the expression 
\begin{equation}
\left[ \frac{- ( \vec{R}_{i}-\vec{R}_{i+1})^{2}  2^{2/\alpha} m^{2/\alpha}}{4 \tau^{2/\alpha} \hbar^{2}}      \right]
\end{equation}
to use as an argument of ${F_\alpha}$. 
From Figure \ref{LogScaleDistribution} above, we know that for large ${x}$,
\begin{equation*}
F_\alpha (x) \sim x^{-P},
\end{equation*} 
for some $P$ that depends on ${\alpha}$.
Thus,
\begin{equation}
\rho_{i}(\tau) \sim (m^{\frac{3N}{\alpha}}) \left(m^{\frac{1}{\alpha}} \sqrt{ (\vec{R}_{i}-\vec{R}_{i+1})^2 }\right)^{-P}.
\end{equation}
As a result, 
\begin{eqnarray}
&& \frac{d\rho_{i}(\tau)}{dm} \sim (m^{\frac{3N}{\alpha}-1}) \left(m^{\frac{1}{\alpha}} \sqrt{ (\vec{R}_{i}-\vec{R}_{i+1})^2}\right)^{-P} \\
&-& P (m^{\frac{3N}{\alpha}}) \left(m^{\frac{1}{\alpha}} \sqrt{ (\vec{R}_{i}-\vec{R}_{i+1})^2}\right)^{-P-1} \sqrt{ (\vec{R}_{i}-\vec{R}_{i+1})^2} \left(m^{\frac{1}{\alpha}-1} \right) \nonumber
\end{eqnarray}
and so
\color{black}
\begin{eqnarray}
\frac{m}{\tau \rho_{i}(\tau)} \frac{d\rho_{i}(\tau)}{dm} &\sim& \frac{m^{-1}}{\tau}
-P \frac{m^{1/\alpha - 1}}{\tau} \left(m^{\frac{1}{\alpha}} \sqrt{ (\vec{R}_{i}-\vec{R}_{i+1})^2}\right)^{-1} \nonumber \\ 
&\text{ }& \hspace{1.5in} \times \sqrt{ (\vec{R}_{i}-\vec{R}_{i+1})^2} \nonumber \\
&=&
\frac{m^{-1}}{\tau}
-P \frac{m^{-1}}{\tau} (m^{-\frac{1}{\alpha}} )
\end{eqnarray}
\color{black}
which is a constant value. 
Thus, ${\langle KE \rangle}$ will initially decay and then approach this value for large $ (\vec{R}_{i}-\vec{R}_{i+1})^2$. 

As may be expected on a conceptual basis and is confirmed in the above plots, the average interbead distance for which the kinetic energy reaches its minimum shifts to larger values with increasing $\alpha$ and $\beta$. This is because larger $\alpha$ imply more diffuse paths, while larger $\beta$ correspond to lower temperatures and therefore larger de Broglie wave lengths. The presence of a minimum in the kinetic energy is a strictly fractional feature which appears for $\alpha$ infinitesimally smaller than two and suggests that fractional particles may exhibit intriguing behavior as they attempt to minimize their potential and kinetic energies -- which have competing minima -- at once.   

\subsection{\label{sec:Free_Particle_Results} Fractional Free Particle PIMC Results}

Before presenting results for $^{4}$He, we first demonstrate our algorithm by simulating the free-particle Hamiltonian. Our interest in the free-particle Hamiltonian stems from the fact that it is amenable to analytic solutions and free-particle PIMC serves as the bedrock for interacting simulations. More specifically, we consider the Hamiltonian

\begin{eqnarray}
\hat{H}_{free,\alpha} = \frac{-\hbar^{\alpha}}{(2m)^{\alpha/2}} \nabla^{\alpha} = -\hbar^{\alpha} D_{\alpha} \nabla^{\alpha} 
\label{Fractional_Hamiltonian_Free}
\end{eqnarray}
for $\alpha \leq 2$ to avoid sampling from negative distributions. In order to make direct comparisons with our helium results presented in the next section, we set our mass equal to that of helium. In both sections, we moreover simulate $N=64$ particles at a density of $\rho=.00323$ bohr$^{-3}$ at all of the temperatures and $\alpha$ values presented.  

\begin{figure}[h]
  \centering
    \includegraphics[width=\columnwidth]{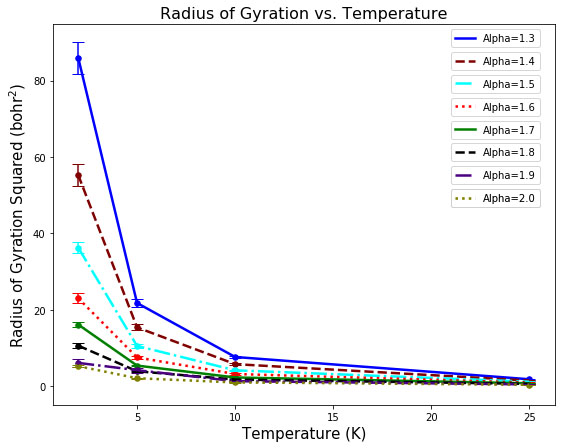}
\caption{The average radius of gyration squared as a function of the temperature for free particles. The radius of gyration squared also steadily increases as $\alpha$ decreases. All results have been converged with respect to the number of beads used at each temperature.}
\label{RG_Noninteracting}
\end{figure}

\begin{figure*}
  \centering
  \includegraphics[width=12cm]{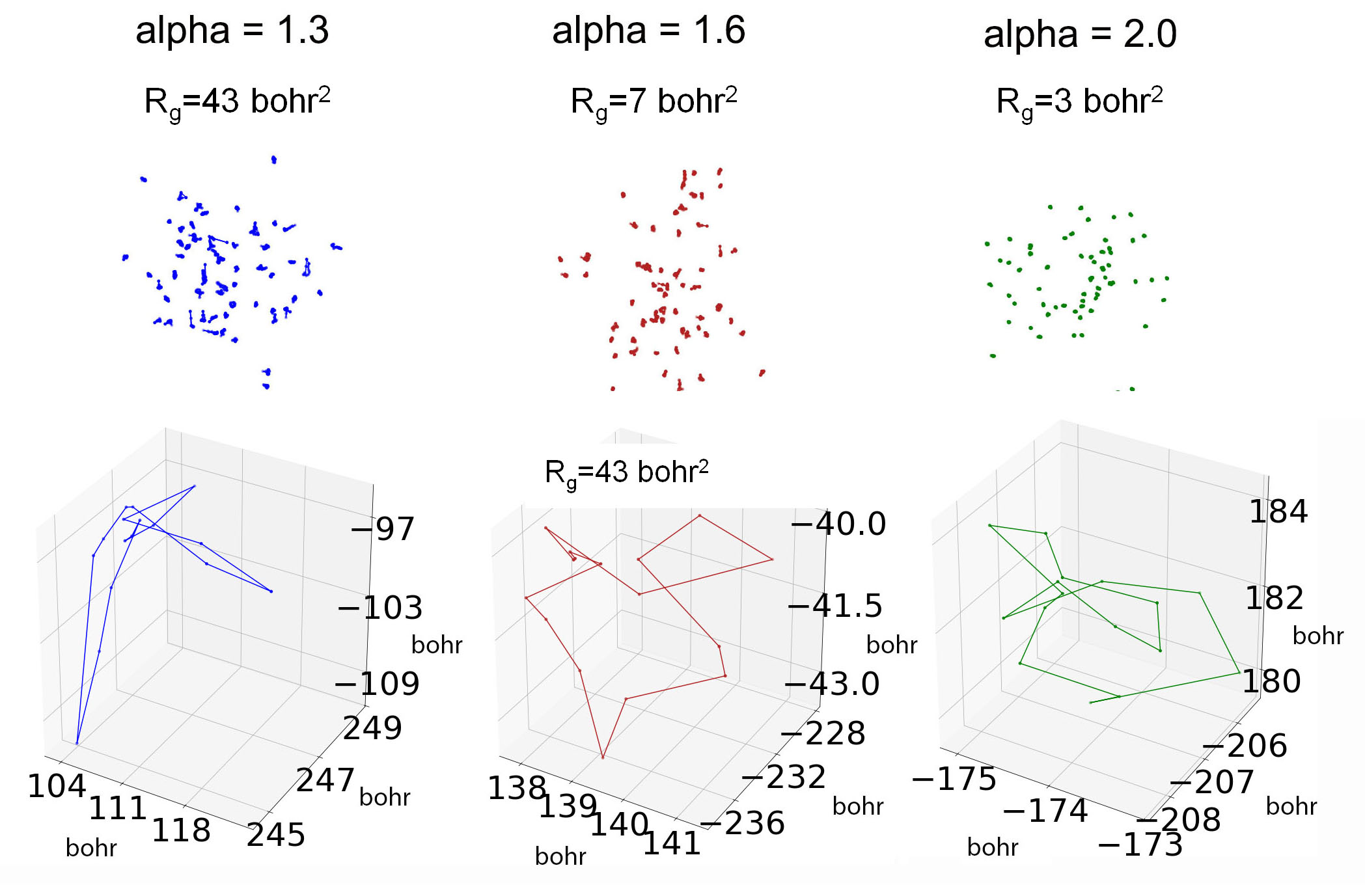}
\caption{Representative paths sampled from the free particle $\alpha=1.3$, $\alpha=1.6$, and $\alpha=2.0$ $T=5$ K distributions. Top: Image of the full $N$=64 particle system after equilibration for the different $\alpha$ values. Bottom: Selected illustrative paths from these systems. Scale units are in bohr. For reference, the radii of gyration squared in bohr$^2$ for the selected paths are 43.13, 7.47, and 2.55, respectively.}
\label{Path_Images}
\end{figure*}

    As a qualitative check on our simulations and underlying theory, we first consider the conformations of the polymers representing our particles. One of the principle motivations for this work was the fact that a fractional Laplacian for $\alpha<2$ should lead to significantly more diffuse particles than in the typical $\alpha=2$ case. As illustrated in Figure \ref{Second_Distribution_Figure}, this is because the $\alpha<2$ fractional density matrix probability distributions possess heavy tails that should permit more frequent sampling of larger interbead distances, $|\vec{R}_{i}-\vec{R}_{i+1}|$. 
\begin{figure}[h]
  \centering
    \includegraphics[width=\columnwidth]{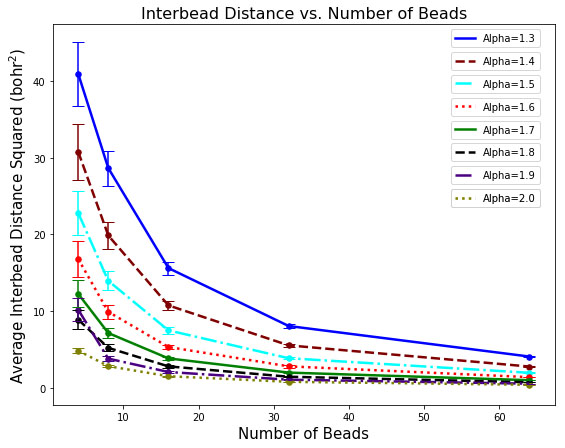}
\caption{The average interbead distance squared as a function of the number of beads at 5K for free particles. The interbead distance steadily increases for the same number of beads as $\alpha$ decreases.}
\label{Interbead_Noninteracting}
\end{figure}        
Figure \ref{Interbead_Noninteracting} demonstrates that this is indeed the case: as $\alpha$ decreases, the average square of the interbead distance, $\langle (\vec{R}_{i}-\vec{R}_{i+1})^{2} \rangle$, for fixed $\beta$ and number of beads steadily increases. As may be expected, this increase is most dramatic at the low temperatures at which, in the absence of a potential, increased delocalization should be favored. Similar trends may be observed in the particles' average radii of gyration, $R_{G} = \langle (\vec{R}_{i} - \vec{R}_{cm})^{2} \rangle$, as depicted in Figure \ref{RG_Noninteracting}. 
Example paths illustrating these features for varying temperatures and fractional exponents are depicted in Figure \ref{Path_Images}. 

What can be gleaned from these plots is that measures of both the average interbead distance and the radius of gyration are accompanied by increasingly large error bars as $\alpha$ decreases. This can be expected as the variance of an $\alpha$-stable law is infinite for $0 < \alpha < 2$, with the mean itself becoming infinite for $\alpha < 1$ (we only consider the $1 < \alpha < 2$ case). Although we have not implemented this in our current code, these variances may be made finite by sampling tempered distributions \cite{Meerschaert_JCOMP} at the cost of approximating the distributions' tails.\cite{Kleinert_Book} 

\begin{figure}[h]
  \centering
    \includegraphics[width=\columnwidth]{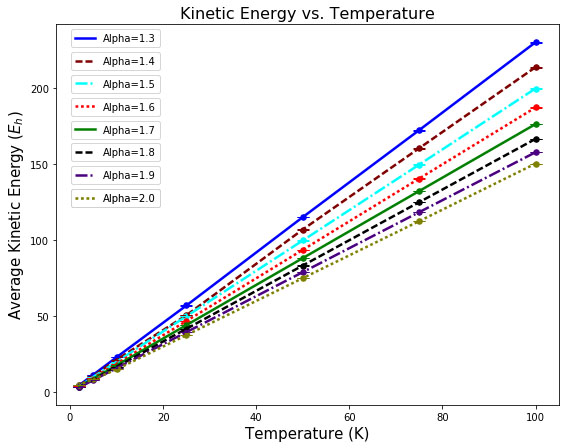}
\caption{The free particle kinetic energy as a function of temperature for varying $\alpha$.}
\label{Kinetic_Non_Interacting}
\end{figure}

Although related to the interbead distance, a more quantitative measure of our fractional paths is the average kinetic energy. In Figure \ref{Kinetic_Non_Interacting}, we plot the simulated PIMC kinetic energy vs. temperature for a range of fractional exponent values. 
As may be expected from the increased interbead distances, the average kinetic energy increases with decreasing $\alpha$. The magnitudes of the kinetic energy may be rationalized by using the fractional version of the equipartition theorem: $\langle KE \rangle = 3 N k_{B} T/\alpha$, which can be readily derived from Equation \eqref{KE_Approx} in the high temperature limit. The larger kinetic energies for smaller $\alpha$ suggests that there is a comparative kinetic energy penalty for fractional cases. While not illustrated here, if the average kinetic energy obtained from PIMC is plotted against the the average interbead distance squared, functional forms possessing minima and plateaus similar to those depicted in Figures \ref{KE_Curves_1} and \ref{KE_Curves_2} are obtained. 

\subsection{\label{sec:Helium_Results} Fractional He-4 PIMC Results}

\begin{figure}[h]
  \centering
  \includegraphics[width=\columnwidth]{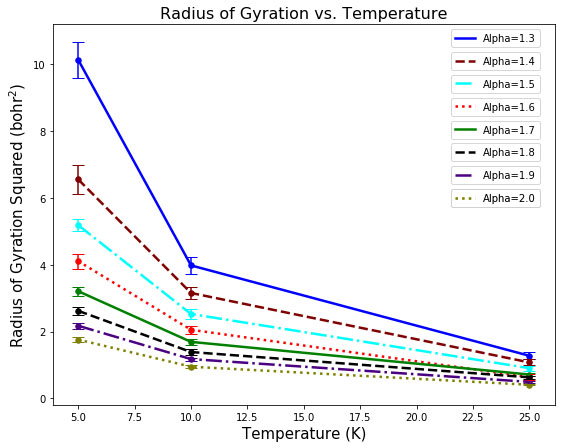}
\caption{The average radius of gyration squared vs. the temperature for $^{4}$He. Note that the radius of gyration squared for $^{4}$He is smaller at all temperatures than it is for free particles. All results have been converged with respect to the number of beads used at each temperature.}
\label{RG_Interacting}
\end{figure}

\begin{figure}[h]
  \centering
  \includegraphics[width=\columnwidth]{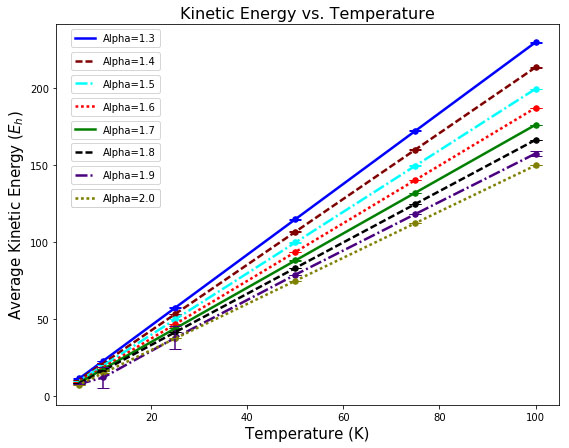}
\caption{The average kinetic energy in Hartree as a function of temperature for $^{4}$He and varying fractional exponents.}
\label{KE_Interacting}
\end{figure}

Building upon our free particle simulations, we also simulate fractional $^{4}$He to demonstrate our PIMC algorithm in the presence of interactions. $^{4}$He is a particularly illuminating example because, even in the $\alpha=2$ case, its particle paths become so diffuse at low temperatures that it can form a superfluid.\cite{Ceperley_Pollock_PRL,Pollock_Ceperley_PRB} In the following, we model $^{4}$He with the Aziz potential.\cite{Aziz_Helium} While the average interacting interbead distance remains similar to the non-interacting interbead distance, the average interacting radius of gyration seems to decrease significantly when compared to the non-interacting radius of gyration. This implies that the interacting particles are more localized (see Figure \ref{RG_Interacting}). This is also borne out in the decreased average kinetic energy depicted in Figure \ref{KE_Interacting}. 
\begin{figure}[h]
  \centering
    \includegraphics[width=\columnwidth]{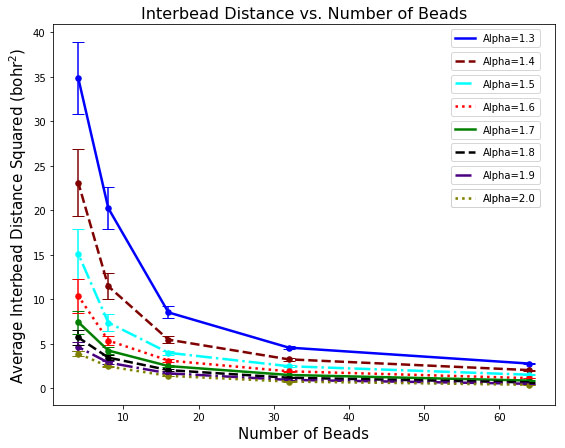}
\caption{The average interbead distance squared vs. the number of beads for $^{4}$He at 5 K.}
\label{Interbead_Distance_Interacting}
\end{figure} 
Irrespective of the increased localization due to the potential, the overarching trend that delocalization is favored with decreasing $\alpha$ persists. 
Interestingly, for $^4$He, we do not observe a pronounced competition between minimizing the kinetic and potential energies. This competition may be more pronounced, leading to more intriguing behavior, for interparticle potentials with deeper minima. 

\section{\label{sec:Conclusions} Summary and Outlook} 

In this work, we have presented a methodology that generalizes the Path Integral Monte Carlo algorithm to fractional Hamiltonians, thereby enabling the computational exploration of fractional Hamiltonians with potentials that are not readily amenable to analytical treatments. We have derived and shown how to sample fractional free particle density matrices by developing a novel approach to sampling Fox-Wright functions. This overall approach may be generalized beyond fractional path integrals to the sampling of density matrices of non-Gaussian forms. We have furthermore employed our algorithm to explore the impact of a fractional Laplacian on free-particle and $^{4}$He path observables, such as the radius of gyration and the average interbead distance. We have demonstrated that the fractional kinetic energy possesses a pronounced minimum as a function of average interbead distance for fractional exponents less than two ($\alpha<2$) and that this minimum shifts to larger interbead distances with decreasing fractional exponents. As such, fractional particles become increasingly more diffuse with decreasing fractional exponents as a consequence of sampling a fractional density matrix distribution with heavier tails than the usual Gaussian distribution. Because the onset of phenomena related to condensation heavily depends on the diffusivity of particle paths, our findings suggest that fractional Hamiltonians may manifest intriguing superfluid and Wigner crystallization phase transitions. Such transitions have only previously been studied for the fractional Schr\"odinger Equation based upon mean field theories.\cite{Kleinert_Paper} Fractional Path Integral Monte Carlo will enable a more complete understanding of such phenomena. 

Before the effects of fractional operators on condensation phenomena may be explored, however, several complications revealed by our work must be resolved. First and foremost, because the conventional staging algorithm\cite{Ceperley_Rev_Mod_Phys,Sprik_Staging} cannot readily be applied to non-Gaussian distributions, we sampled our partition functions using highly inefficient single bead moves in this work. While, for the potentials explored, we verified that our simulations were able to converge to their equilibrium distributions, we would not expect this to be the case for more general potentials. It is furthermore easy to imagine that single bead moves would be suboptimal for sampling permutations among particles with quantum statistics. Future work therefore necessitates the development of generalized multi-bead moves. Fourier path sampling may permit the generality we require.\cite{Doll_Fourier_1,Doll_Fourier_2,Krauth_Statistical}

As Figures \ref{First_Distribution_Figure} and \ref{Second_Distribution_Figure} illustrate, the density matrix distributions sampled for $x-x'$ in our algorithm are furthermore significantly more complex than the Gaussian distribution sampled conventionally. For $\alpha<2$, these distributions possess heavy tails, while for $\alpha>2$, these distributions can be negative. The discrete inverse-transform sampling method we presented can accurately sample the heavy tails, but there remain theoretical questions about the effect of heavy tails in the theory. Namely, whether the infinite variance (for $\alpha<2$) and infinite mean (for $\alpha\le1$) of the fractional density matrix requires the use of other notions of central tendency (such as median) for observables. This is a fascinating question for future study. In this direction, it may be fruitful to explore ``tempered-fractional quantum mechanics,'' which has not been studied to our knowledge. This would conceivably employ tempered Levy distributions instead of pure heavy-tailed distributions, and temperated fractional derivatives instead of standard fractional derivatives.\cite{Meerschaert_JCOMP} Tempered distributions enjoy the heavy tail to a certain argument, but then decay exponentially to avoid issues with infinite variance. 

\color{black}
It may also be of interest to consider the physics of the time-fractional Schrodinger Equation,\cite{Naber_Time} and even the space-and-time-fractional Schrodinger Equation.\cite{Wang_Xu_Time} These equations introduce waiting times in addition to long jumps. However, unlike the space-fractional equation, the properties of the time-fractional Schrodinger equation suggest that it cannot be used as a probabalistic theory of Quantum Mechanics.\cite{Dong_Xu_Time, Laskin_Time} Thus, more work is needed before any implementation of a time-fractional model is justified.
\color{black}

In summary, we have proposed a novel fractional Path Integral Monte Carlo algorithm. Our algorithm provides a clear numerical path toward unraveling the fractional physics of interacting systems, much of which could only be approximated analytically in the past. We look forward to employing this algorithm to explore where our fully interacting predictions differ from previous approximate analytical solutions and to the study of such novel phenomena as fractional superfluidity. 

\acknowledgements

The authors would like to dedicate this work to the late Jimmie Doll. Jimmie initially convened the authors to brainstorm about this topic, but unfortunately, did not live to see this work to its fruition. The authors would also like to thank George Karniadakis, who suggested to M.G. that he explore this topic and contact Jimmie. Finally, we thank Paul Dupuis, Matthew Harrison, and Mark Meerschaert for valuable discussions. 

M. G. would like to acknowledge support from the OSD/ARO/MURI grant ``Fractional PDEs for Conservation Laws and Beyond: Theory, Numerics and Applications (W911NF-15-1-0562).'' Part of this research was conducted using computational resources and services at the Center for Computation and Visualization, Brown University. 

\section{\label{Appendix_1} Appendix I: Separability of the Fractional Density Matrix in Multiple Dimensions}

In this work, we have derived and demonstrated how to sample the one-dimensional form of the free fractional density matrix. Here, we reaffirm that, just as in the $\alpha = 2$ case, a multi-dimensional free density matrix is simply a product of one-dimensional free density matrices. 

Recalling Equation \eqref{ThreeD_DensityMatrix}, the free multi-dimensional density matrix may be written as

\begin{eqnarray}
\rho(\vec{R}, \vec{R}^{'}; \tau) &=& \langle \vec{R}| e^{-\tau \hat{K}} | \vec{R}^{'} \rangle \nonumber \\ &=&
\sum_{j} \phi_{k,j}^{*}(\vec{R}) \phi_{k,j}(\vec{R}^{'}) e^{-\tau E_{k,j}}, 
\label{ThreeD_DensityMatrix_2}
\end{eqnarray}
where $\phi_{k,j}(\vec{R})$ denote the eigenvectors and $E_{k,j}$ denote the eigenvalues of the Hamiltonian, which in this case, reduces to the kinetic operator. In Equation \eqref{ThreeD_DensityMatrix_2}, $k$ denotes that these are kinetic eigenvectors/eigenvalues, while $j$ represents the eigenvector/eigenvalue index. Assuming particle masses are the same, the many body, multi-dimensional Laplacian in Equation \eqref{Kinetic_Operator} may be expanded into 

\begin{equation}
\hat{K}  = - \lambda \nabla^{\alpha} = -\lambda \sum_{i}^{N} \left( \nabla_{i,x}^{\alpha} + \nabla_{i,y}^{\alpha} + \nabla_{i,z}^{\alpha} \right), 
\end{equation}
where $i$ denotes the particle number. Since the differential operators act on different particles in orthogonal directions, they commute with one another. Moreover, because the Hamiltonian is non-interacting, the many body wave function is separable

\begin{equation}
|\vec{R} \rangle = |x_{1} \rangle \otimes | y_{1} \rangle \otimes | z_{1} \rangle \otimes ...\otimes |x_{N} \rangle \otimes | y_{N} \rangle \otimes | z_{N} \rangle.
\end{equation}
This implies that 

\begin{eqnarray}
&& \rho(\vec{R}, \vec{R}^{'}; \tau) = \langle \vec{R}| e^{-\tau \hat{K} } | \vec{R}^{'} \rangle \nonumber \\ &=&
\prod_{i}^{N} \langle x_{i} | e^{\tau \lambda \nabla_{i,x}^{\alpha}} | x_{i}' \rangle  \langle y_{i} | e^{\tau \lambda \nabla_{i,y}^{\alpha}} | y_{i}' \rangle \langle z_{i} | e^{\tau \lambda \nabla_{i,z}^{\alpha}} | z_{i}' \rangle  \nonumber \\ &=& 
\prod_{i}^{N} \rho(x_{i}, x_{i}'; \tau)\rho(y_{i}, y_{i}'; \tau) \rho(z_{i}, z_{i}'; \tau),
\label{Density_Matrix_Breakup}
\end{eqnarray}
where $\rho(x_{i}, x_{i}', \tau)$ denotes a uni-dimensional density matrix of the same form as given by Equations \eqref{Initial_Fractional_Density_Matrix} and \eqref{distribution}. Sampling the many body, multi-dimensional density matrix thus reduces to independently sampling the one-dimensional density matrices for all $x_{i}$, $y_{i}$, and $z_{i}$ coordinates. 
\section{\label{Appendix_2} Appendix II: Suzuki-Trotter Factorization Using Fractional Laplacians}
In order to use the Path Integral Monte Carlo algorithm for the fractional case, it is essential to first demonstrate that the fractional density matrices obey the density matrix convolution principle given by Equation \eqref{Convolution_Equation}. Starting with the one-dimensional expression for the kinetic density matrix for simplicity, 
\begin{equation}
\rho^{per,L}(x,x',\beta) = \frac{1}{2\pi} \int_{-\infty}^{\infty} dC e^{iC(x-x')}e^{-\beta D_{\sigma} \hbar^{\alpha} |C|^{\alpha}}
\end{equation}
this factorization may be verified via substitution

\begin{eqnarray} 
&& \rho^{per,L}(x,x',\beta) \nonumber \\ &=&  \int_{-\infty}^{\infty} dx'' \rho^{per,L}(x,x'',\beta/2) \rho^{per,L}(x'',x',\beta/2) \nonumber \\ &=& \frac{1}{2\pi} \int_{-\infty}^{\infty} dx'' \int_{-\infty}^{\infty} dC e^{iC(x-x'')} e^{-(\beta/2) D_{\sigma} \hbar^{\alpha}|C|^{\alpha}} \nonumber \\ &&
\frac{1}{2\pi} \int_{-\infty}^{\infty} dC' e^{iC'(x''-x')} e^{-(\beta/2) D_{\sigma} \hbar^{\alpha} |C'|^{\alpha}} \nonumber \\
&=& \frac{1}{4\pi^{2}} \int_{-\infty}^{\infty} \int_{-\infty}^{\infty} dC dC' e^{iCx} e^{-iCx'} \nonumber \\ &&
\left[\int_{-\infty}^{\infty} dx'' e^{ix''(C'-C)} \right] e^{-(\beta/2) D_{\sigma} \hbar^{\alpha} |C|^{\alpha}} e^{-(\beta/2) D_{\sigma} \hbar^{\alpha} |C'|^{\alpha}} \nonumber  \\
&=& \frac{1}{4\pi^{2}} \int_{-\infty}^{\infty} \int_{-\infty}^{\infty}  dC dC' e^{iCx} e^{-iCx'} \left[2 \pi \delta(C-C') \right] \nonumber \\ &&e^{-(\beta/2) D_{\sigma} \hbar^{\alpha} |C|^{\alpha}} 
e^{-(\beta/2) D_{\sigma} \hbar^{\alpha} |C'|^{\alpha}} \nonumber  \\
&=& \frac{1}{2\pi} \int_{-\infty}^{\infty} dC e^{iC(x-x')} e^{-(\beta/2) D_{\sigma} \hbar^{\alpha} |C|^{\alpha}} e^{-\beta/2 D_{\sigma} \hbar^{\alpha} |C|^{\alpha}} \nonumber \\
&=& \frac{1}{2\pi} \int_{-\infty}^{\infty} dC e^{iC(x-x')} e^{-\beta D_{\sigma} \hbar^{\alpha} |C|^{\alpha}} \nonumber \\
&=& \rho^{per,L}(x,x'),
\end{eqnarray}
where the substitution $e^{ix''(C'-C)} = 2\pi \delta(C'-C)$ was made. This demonstrates that the fractional kinetic density matrix at a large imaginary time, $\beta$, may be factored into a convolution over short imaginary time density matrices. 
 
In order to factor the full short time propagators into kinetic and potential propagators, it must moreover be verified that Suzuki-Trotter factorization\cite{Trotter_1959} can be performed on the fractional Laplacian. As discussed by Simon,\cite{Simon_1979} Suzuki-Trotter factorization is valid as long as the kinetic and potential operators are self-adjoint\cite{Laskin_4} and bounded from below, which holds by the definition of the (negative) fractional Laplacian in $\mathbb{R}^n$ as the power, in the sense of the spectral theory, of the Laplacian. This results in a self-adjoint operator\cite{Rudin_Functional} with positive eigenvalues.\cite{Hermann_Fractional} Because the density matrix convolution property and the Trotter product formula hold, we can factor for $\alpha<2$ cases, as usual.   
 
\section{\label{Appendix_3} Appendix III: Derivation of the Expression for the Fractional Kinetic and Total Energies}

The thermodynamic estimator for the kinetic energy is the mass derivative of the partition function

\begin{equation}
\langle \hat{K} \rangle = \frac{m}{\beta} \frac{\partial Z}{\partial m}. 
\end{equation}
Following Ceperley,\cite{Ceperley_Rev_Mod_Phys} all $\vec{R}_{i}$-$\vec{R}_{i+1}$ links may be viewed as equivalent and we may therefore take the derivative of the single link density matrix, as opposed to the full partition function, and average over all links. Neglecting the potential contribution to the Hamiltonian for now, let 

\begin{widetext}
\begin{eqnarray}
\rho_{i}(\tau) \equiv \rho(\vec{R}_{i}, \vec{R}_{i+1}; \tau) &=& \left( \frac{ 2^{1/\alpha} m^{1/\alpha}}{\sqrt{\pi}\tau^{1/\alpha}\hbar \alpha} \right)^{3 N} \left[ \sum_{n=0}^{\infty} \frac{\Gamma(1/\alpha + 2/\alpha n)}{\Gamma(1/2+n)} \left[ \frac{- ( \vec{R}_{i}-\vec{R}_{i+1})^{2}  2^{2/\alpha} m^{2/\alpha}}{4 \tau^{2/\alpha} \hbar^{2}}      \right]^{n}/n! \right].
\end{eqnarray}
\end{widetext}

Here, it should be noted that the $\vec{R}_{i}$ and $\vec{R}_{i+1}$ beads are meant to be adjacent and the corresponding vectors will be replaced by $\vec{R}$ and $\vec{R}'$ in what follows. 

\begin{widetext}
\begin{eqnarray}
&& \langle KE \rangle_{link} = \left \langle \frac{m}{\tau \rho_{i}(\tau)}\frac{\partial \rho_{i}(\tau)}{\partial m} \right \rangle_{link} \nonumber \\
&=& \left \langle \frac{m}{\tau \rho_{i}(\tau)} \frac{3 N}{\alpha m} \left( \frac{2^{1/\alpha} m^{1/\alpha}}{\sqrt{\pi} \tau^{1/\alpha} \hbar \alpha} \right)^{3N} 
\left[  \sum_{n=0}^{\infty} \frac{ \Gamma(1/\alpha + 2/\alpha n)}{\Gamma(1/2 + n)} \left[ \frac{-(\vec{R}-\vec{R'})^{2}  2^{2/\alpha} m^{2/\alpha}}{4 \tau^{2/\alpha} \hbar^{2}}      \right]^{n}/n! \right] \right \rangle_{link} \nonumber \\ 
&+& \left \langle \frac{m}{\tau \rho_{i}(\tau)} \left( \frac{ 2^{1/\alpha} m^{1/\alpha}}{\sqrt{\pi}\tau^{1/\alpha}\hbar \alpha} \right)^{3 N}   
 \left[ \sum_{n=1}^{\infty} \frac{\Gamma(1/\alpha + 2/\alpha n)}{\Gamma(1/2+n)} \frac{2n}{\alpha m} \left[ \frac{- (\vec{R}-\vec{R'})^{2} 2^{2/\alpha} m^{2/\alpha}}{4 \tau^{2/\alpha} \hbar^{2}}      \right]^{n}/n! \right] \right \rangle_{link} \nonumber  \\
&=& \left \langle \frac{m}{\tau} \frac{3N}{\alpha m} \right \rangle_{link} \nonumber \\ 
&+& \left \langle \frac{m}{\tau}\frac{2}{\alpha m} \left[ \sum_{n=1}^{\infty} \frac{\Gamma(1/\alpha + 2/\alpha n)}{\Gamma(1/2+n)} n \left[ \frac{-(\vec{R}-\vec{R'})^{2} 2^{2/\alpha} m^{2/\alpha}}{4 \tau^{2/\alpha} \hbar^{2}}      \right]^{n}/n! \right]/ \left[ \sum_{n=0}^{\infty} \frac{\Gamma(1/\alpha + 2/\alpha n)}{\Gamma(1/2+n)}  \left[ \frac{- (\vec{R}-\vec{R'})^{2}  2^{2/\alpha} m^{2/\alpha}}{4 \tau^{2/\alpha} \hbar^{2}}      \right]^{n}/n! \right] \right \rangle_{link} \nonumber \\
&=& \left \langle \frac{m}{\tau} \frac{3N}{\alpha m} \right \rangle_{link} + \left \langle \frac{m}{\tau}\frac{2}{\alpha m} \left[ \sum_{n=0}^{\infty} \frac{\Gamma(1/\alpha + 2/\alpha (n+1))}{\Gamma(1/2+(n))}  \left[ \frac{- (\vec{R}-\vec{R'})^{2}  2^{2/\alpha} m^{2/\alpha}}{4 \tau^{2/\alpha} \hbar^{2}}      \right]^{n+1}/n! \right]/ \right. \nonumber \\
&& \left. \left[ \sum_{n=0}^{\infty} \frac{\Gamma(1/\alpha + 2/\alpha n)}{\Gamma(1/2+n)}  \left[ \frac{- (\vec{R}-\vec{R'})^{2}  2^{2/\alpha} m^{2/\alpha}}{4 \tau^{2/\alpha} \hbar^{2}}      \right]^{n}/n! \right] \right \rangle_{link} \nonumber \\
&=& \left \langle  \frac{m}{\tau} \frac{3N}{\alpha m} \right \rangle_{link}  \left \langle \frac{m}{\tau}\frac{2}{\alpha m} \left[\frac{- (\vec{R}-\vec{R'})^{2}  2^{2/\alpha} m^{2/\alpha}}{4 \tau^{2/\alpha} \hbar^{2}}      \right] \left[ \sum_{n=0}^{\infty} \frac{\Gamma(1/\alpha + 2/\alpha (n+1))}{\Gamma(1/2+(n+1))}  \left[ \frac{- (\vec{R}-\vec{R'})^{2}  2^{2/\alpha} m^{2/\alpha}}{4 \tau^{2/\alpha} \hbar^{2}}      \right]^{n}/n! \right]/ \right. \nonumber \\
&& \left. \left[ \sum_{n=0}^{\infty} \frac{\Gamma(1/\alpha + 2/\alpha n)}{\Gamma(1/2+n)}  \left[ \frac{- (\vec{R}-\vec{R'})^{2}  2^{2/\alpha} m^{2/\alpha}}{4 \tau^{2/\alpha} \hbar^{2}}      \right]^{n}/n! \right] \right \rangle_{link} \nonumber  \\
&=& \left \langle \frac{3N}{\tau \alpha} \right \rangle_{link}  \left \langle \frac{2}{\alpha \tau} \left[\frac{- (\vec{R}-\vec{R'})^{2}   2^{2/\alpha} m^{2/\alpha}}{4 \tau^{2/\alpha} \hbar^{2}}      \right]\left[ \sum_{n=0}^{\infty} \frac{\Gamma(1/\alpha + 2/\alpha (n+1))}{\Gamma(1/2+(n+1))}  \left[ \frac{- (\vec{R}-\vec{R'})^{2}  2^{2/\alpha} m^{2/\alpha}}{4 \tau^{2/\alpha} \hbar^{2}}      \right]^{n}/n! \right]/ \right. \nonumber \\
&& \left. \left[ \sum_{n=0}^{\infty} \frac{\Gamma(1/\alpha + 2/\alpha n)}{\Gamma(1/2+n)}  \left[ \frac{- (\vec{R}-\vec{R'})^{2}  2^{2/\alpha} m^{2/\alpha}}{4 \tau^{2/\alpha} \hbar^{2}}      \right]^{n}/n! \right] \right \rangle_{link} \nonumber \\
\label{Derivative_KE}
\end{eqnarray}
\end{widetext}
In the above, the average $\langle \rangle_{link}$ denotes an average over all links within the polymers. As a check, this simplifies to the usual expression for the kinetic energy upon substitution of $\alpha=2$ 

\begin{eqnarray}
\langle KE \rangle_{link} &=& \frac{3 N}{2 \tau} - \frac{ m \langle (\vec{R} - \vec{R}')^{2} \rangle_{link} }{2 \tau^{2}\hbar^{2}}. 
\end{eqnarray}

Because Equation \eqref{Derivative_KE} contains a factor of $n+1$ in its gamma functions, this makes its related series more difficult to sum to convergence. We have likewise resorted to using a three-point forward difference formula in which each density matrix is evaluated at multiple values of the mass. When the density matrices become exceedingly large, it is often more numerically tractable to make the replacement    

\color{black}
\begin{equation}
\frac{m}{\tau \rho_{i}(\tau)} \frac{ \partial \rho_{i}(\tau)}{\partial m} = \frac{m}{\tau} \frac{ \partial \ln \rho_{i}(\tau)}{\partial m}, 
\end{equation}
\color{black}
and to therefore take the derivative of the \color{black}$\ln$ \color{black} of the partition function instead. 

\bibliography{ref}{}

\end{document}